\documentclass[tightenlines,eqsecnum,floats,aps,nofootinbib,showpacs,12pt]{iopart}

\usepackage[dvips]{color}
\usepackage{graphicx}
\usepackage{graphicx}
\usepackage{enumerate} 
\usepackage{colordvi} 
\usepackage{bm}

\newcommand{\simgt}{\lower.5ex\hbox{$\; \buildrel > \over \sim \;$}}
\newcommand{\simlt}{\lower.5ex\hbox{$\; \buildrel < \over \sim \;$}}
\newcommand{\fnl}{f_{\rm NL}}
\newcommand{\gnl}{g_{\rm NL}}
\newcommand{\bfx}{\mbox{\boldmath$x$}}
\newcommand{\bfk}{\mbox{\boldmath$k$}}
\newcommand{\bfp}{\mbox{\boldmath$p$}}
\newcommand{\bfq}{\mbox{\boldmath$q$}}
\newcommand{\cyc}{({\rm cyc.})}
\newcommand{\deltac}{\delta_{\rm c}}
\newcommand{\deltag}{\delta_{\rm g}}

\newcommand{\deltam}{\delta_{\rm m}}
\newcommand{\sigmar}{\sigma_{\rm R}}
\newcommand{\Pm}{P_{\rm m}}
\newcommand{\Pg}{P_{\rm g}}

\newcommand{\Bm}{B_{\rm m}}
\newcommand{\Bg}{B_{\rm g}}
\newcommand{\Bh}{B_{\rm h}}
\newcommand{\Tm}{T_{\rm m}}
\newcommand{\Omegam}{\Omega_{\rm m}}
\newcommand{\Wr}{\tilde{W}_R}
\newcommand{\Mr}{\mathcal{M}_R}
\newcommand{\Fr}{\mathcal{F}_R}
\newcommand{\Mm}{\mathcal{M}}

\begin{document}

\title[]
{Scale Dependence of Halo Bispectrum from Non-Gaussian Initial Conditions 
  in Cosmological N-body Simulations}

\author{Takahiro Nishimichi$^{1,3}$, Atsushi Taruya$^{2,3}$, Kazuya
  Koyama$^4$, Cristiano Sabiu$^{4,5}$} 
\address{$^1$~Department of
  Physics, School of Science, The University of Tokyo, Bunkyo-ku,
  Tokyo 113-0033, Japan} 
\address{$^2$~Research Center for Early
  Universe, School of Science, The University of Tokyo, Bunkyo-ku,
  Tokyo 113-0033, Japan} 
\address{$^3$~Institute for the Physics and
  Mathematics of the Universe, University of Tokyo, Kashiwa, Chiba
  277-8568, Japan} 
\address{$^4$~Institute of Cosmology and
  Gravitation, Dennis Sciama Building, University of Portsmouth,
  Portsmouth, PO1 3FX, UK} 
\address{$^5$~Department of Physics \&
  Astronomy, University College London, Gower Street, London, UK}

\ead{takahiro.nishimichi@ipmu.jp}

\begin{abstract}
    We study the halo bispectrum from
    non-Gaussian initial conditions. Based on a set of large $N$-body
    simulations starting from initial density fields with local type
    non-Gaussianity, we find that the halo bispectrum exhibits a
    strong dependence on the shape and scale of Fourier space triangles
    near squeezed configurations at large scales. The amplitude of 
    the halo bispectrum roughly scales as $\fnl^2$. The resultant scaling 
    on the triangular shape is consistent with that predicted by Jeong \& Komatsu
    based on perturbation theory. We systematically investigate this dependence 
    with varying redshifts and halo mass thresholds. 
    It is shown that the $\fnl$ dependence of the halo bispectrum is stronger for more
    massive haloes at higher redshifts.
     This feature can be a useful
    discriminator of inflation scenarios in future deep and wide
    galaxy redshift surveys.
\end{abstract}

\maketitle

\section{Introduction}
The standard cosmological model has successfully explained the
observed statistical properties of the cosmic microwave background
(CMB) radiation and the large scale structure (LSS) traced by galaxies
(e.g., \cite{WMAP5,Tegmark2006}). The model usually assumes that the
primordial density/temperature/curvature fluctuations follow Gaussian
statistics. Recently, however, possible deviations from standard
Gaussian statistics has attracted great attention with rapid
progress in observational techniques. It offers an opportunity to
access cosmological information beyond traditional power spectrum
analysis. Many recent works have discussed the constraints and future
detectability of possible deviations from Gaussianity through the
observations of CMB and LSS (e.g., \cite{Komatsu2001,Carbone2008}).

According to the inflationary scenarios, primordial curvature
perturbations are generated during the accelerated phase of cosmic
expansion.  The simplest inflation models, in which the inflation
takes place with the slow-roll single scalar field that has a
canonical kinetic structure, generally predicts a nearly
scale-invariant spectrum of curvature perturbations, and a small
departure from Gaussianity.  On the other hand, a variety of inflation
models that produce a large non-Gaussianity has been recently proposed
(see \cite{Bartolo2004} for a review).  Among these, the models with
large non-Gaussianity generated by non-linear dynamics on
super-horizon scales can have a generic prediction for non-Gaussian
properties.  Denoting a Gaussian field by $\Phi_{\rm G}(\bfx)$, the
Bardeen's curvature perturbation during the matter era is
characterized as \cite{Komatsu2001}:
\begin{eqnarray}
\Phi(\bfx) = \Phi_{\rm G}(\bfx)+\fnl\left\{
\Phi_{\rm G}^2(\bfx)-\langle\Phi_{\rm G}^2(\bfx)\rangle\right\}
+\cdots.
\label{eq:fnllocal}
\end{eqnarray}
This type of non-Gaussianity, described as a local function of
the Gaussian field, is often called {\it local type} non-Gaussianity, and
the leading-order coefficient $\fnl$, which controls the strength of
non-Gaussianity, has information on the generation mechanisms for
non-Gaussian fluctuations.  Although the current constraint on the
parameter $\fnl$ from CMB data is $-9<\fnl<111$ ($95\%$C.L.)
\cite{WMAP5} and it is still consistent with Gaussian ($\fnl=0$), the
constraint will be tightened by the on-going CMB experiment, Planck
\cite{Planck}. As standard inflation models generally predict
$|\fnl|\ll1$, detection of large non-Gaussianity immediately implies
the non-standard mechanism for generation of primordial curvature
perturbations.

In this paper, we focus on how this type of non-Gaussianity alters the
statistical properties of LSS.  Let us first define the power spectrum
of mass density fluctuations assuming isotropy and homogeneity:
\begin{eqnarray}
\langle \deltam(\bfk_1)\deltam(\bfk_2)\rangle \equiv (2\pi)^3\delta_{\rm D}(\bfk_1+\bfk_2)\Pm(k_1),
\label{eq:pow_def}
\end{eqnarray}
where $\deltam(\bfk)$ denotes the Fourier transform of the density
contrast, while $\delta_{\rm D}(\bfk)$ represents the Dirac delta
function. If the density field follows Gaussian statistics, its power
spectrum determines all the statistical properties. Next we define the
bispectrum of a mass density field:
\begin{eqnarray}
\langle \deltam(\bfk_1)\deltam(\bfk_2)\deltam(\bfk_3)\rangle \equiv (2\pi)^3\delta_{\rm D}
(\bfk_1+\bfk_2+\bfk_3)\Bm(\bfk_1,\bfk_2,\bfk_3).
\label{eq:bis_def}
\end{eqnarray}
Since this is the lowest-order non-vanishing quantity in the presence
of non-Gaussianity, the bispectrum is naively expected as the best
measure for non-Gaussianity (e.g.,
\cite{Scoccimarro2000,Verde2000,Scoccimarro2004,Sefusatti2007}).

Recently, however, the {\it galaxy} or {\it halo} power spectrum has
been reconsidered in the presence of local and/or equilateral type
primordial non-Gaussianity both analytically and numerically (e.g.,
\cite{Dalal2008,Afshordi2008,Taruya2008,Matarrese2008,Pillepich2008,Desjacques2009a,Desjacques2009b,Grossi2009,Slosar2008,McDonald2008,Sefusatti2009,Giannantonio2009}).
The matter power spectrum or bispectrum is not a direct observable,
and the real measurement of LSS gives {\it galaxy} power spectrum or
bispectrum, as defined similarly to equations (\ref{eq:pow_def}) and
(\ref{eq:bis_def}).  Since galaxies are biased tracers of the dark
matter distribution, the information of $\fnl$ is imprinted in a
different manner: a new contribution coming from the primordial
non-Gaussianity may dominate over the Gaussian term in the galaxy
power spectrum, $\Pg(k)$, at very large scales
($k\simlt0.01h$Mpc$^{-1}$).  This new contribution, which is sometimes
referred to as the ``scale-dependent bias'' of the galaxy power
spectrum, may be a powerful indicator to constrain $\fnl$.  Indeed, it
has been recently applied to the clustering statistics of SDSS LRG and
quasar samples, and the tight constraints on $\fnl$ are comparable to
those obtained from CMB measurements have been obtained
\cite{Slosar2008}.

The purpose of this paper is to examine the bispectrum of biased
tracers in detail. While the matter bispectrum in the presence
of primordial non-Gaussianity has been studied in the literature using
both perturbation theory and numerical simulations, the galaxy
bispectrum may significantly differ from the matter bispectrum in the
presence of primordial non-Gaussianity, just like the difference in
the power spectra.  Since the local type non-Gaussianity can be
straightforwardly implemented within $N$-body simulations, numerical
study on the bispectrum for the dark matter haloes is the first
important step toward a practical understanding of the galaxy
bispectrum.

Incidentally, Jeong and Komatsu (2009) recently proposed a new
parametrized model for the halo/galaxy bispectrum (\cite{Jeong2009}, see
also \cite{Sefusatti2009}) based on the peak bias model
\cite{Matarrese1986} and the local bias model \cite{Fry1993}. They
found that the formula for the galaxy bispectrum used in
\cite{Sefusatti2007} was missing important contributions from the
scale dependent bias effects and they discovered new terms that are
important at ``squeezed'' configurations where $k_1, k_2\gg k_3$. It
was argued that these new contributions enable us to put stronger
constraints on $f_{\rm NL}$ than those obtained in
\cite{Sefusatti2007}.  It is of great importance to confirm the
scale-dependent bias effects in the bispectrum by $N$-body
simulations.

The rest of this paper is organized as follows: we first review the
analytical models of the power spectrum and the bispectrum of biased
tracers in section \ref{sec:theory}. We then describe the setup and
initial conditions for $N$-body simulations in section
\ref{sec:simulation}. As a first check of our simulations, in section
\ref{sec:pow_results}, we compute the matter and halo power spectra,
and the results are compared with previous works.  Section
\ref{sec:bis_results} gives the main results of this paper, in which
the simulation results for the matter and halo bispectra are presented
and compared with predictions from analytic models, particularly focusing
on their scale dependence.  The dependence of the halo bispectrum on
the halo mass threshold and redshift is also investigated in detail.
Section \ref{sec:discussion} discusses the future prospects for
detecting the primordial non-Gaussianity using the scale-dependent
properties of the halo/galaxy bispectrum. Finally, section
\ref{sec:conclusion} is devoted to conclusions and discussions.

\section{Theoretical models}
\label{sec:theory}

In this section, we summarize the theoretical predictions of the power
spectrum and bispectrum.  We use perturbation theory to examine the
matter power spectrum and bispectrum, and then present those of biased
tracers based on the local bias model. For the scales of our interest
($k\simlt0.1h$ Mpc$^{-1}$), the non-linearity of gravitational
evolution is moderate and the perturbation theory is valid and
trustful.  We especially focus on the behavior of the bispectrum in
the squeezed limit on large scales, where $k\equiv k_1 = k_2 \equiv
\alpha k_3$, $k\to0$ and $\alpha\gg1$.

Let us first consider the matter density fluctuation.  We
perturbatively expand this as
\begin{eqnarray}
\deltam(\bfk;z) = \deltam^{(1)}(\bfk;z)+\deltam^{(2)}(\bfk;z)+\deltam^{(3)}(\bfk;z)+\cdots.
\end{eqnarray}
The linear-order solution is related to the Bardeen's curvature
perturbation in equation (\ref{eq:fnllocal}) in Fourier space by
\begin{eqnarray}
\deltam^{(1)}(\bfk;z) = {\cal M}(k;z)\Phi(\bfk),
\end{eqnarray}
where the conversion factor $\mathcal{M}$ is defined as
\begin{eqnarray}
{\cal M}(k;z) \equiv \frac23\frac{k^2T(k)D(z)}{\Omegam H_0^2}.
\label{eq:M}
\end{eqnarray}
In the above, $\Omegam$ is the current matter density normalized by
the critical density, $H_0$ is current Hubble constant, $T(k)$ denotes
the matter transfer function normalized to unity at $k\to0$ and $D(z)$
is the linear growth rate normalized to the scale factor in the limit
of matter dominant era.  The higher-order solutions,
$\deltam^{(n)}(\bfk;z)$, are formally written as
\begin{eqnarray}
\hspace{-2cm}\deltam^{(n)}(\bfk;z) = \int d^3\bfq_1\cdots d^3\bfq_n\delta_{\rm D}(\bfk-\bfq_{1\dots n})
F_n(\bfq_1,\dots,\bfq_n)\deltam^{(1)}(\bfq_1;z)\cdots\deltam^{(1)}(\bfq_n;z),
\end{eqnarray}
where $\bfq_{1\dots n}\equiv\bfq_1+\cdots+\bfq_n$, and $F_n$ are the
kernel functions (see \cite{Bernardeau2002} for a review).  Then
keeping terms up to fourth order in $\deltam^{(1)}$, the power
spectrum and bispectrum of the matter density fluctuations are given
by (e.g., \cite{Taruya2008,Sefusatti2007})
\begin{eqnarray}
    \hspace{-1cm}\Pm(k;z) &=& P_0(k;z)\;+\;2\int \frac{d^3\bfq}{(2\pi)^3}F_2(\bfq,\bfk-\bfq)
    B_0(-\bfk,\bfq,\bfk-\bfq;z)
    \label{eq:pow_pt}\nonumber\\
    &&+\;2\int \frac{d^3\bfq}{(2\pi)^3}\left\{F_2(\bfq,\bfk-\bfq)\right\}^2P_0(q;z)P_0(|\bfk-\bfq|;z)
    \nonumber\\
    &&+\;\int \frac{d^3\bfp d^3\bfq}{(2\pi)^6}F_2(\bfp,\bfk-\bfp)F_2(\bfq,-\bfk-\bfq)
    T_0(\bfp,\bfk-\bfp,\bfq,-\bfk-\bfq;z),\nonumber\\
    &&+P_0(k;z)\int\frac{\rmd^3\bfq}{(2\pi)^3}F_3(\bfk,\bfq,-\bfq)P_0(q;z),\\
    \hspace{-2.4cm}\Bm(k_1,k_2,k_3;z) &=& 2\fnl\left[\frac{P_0(k_1;z)P_0(k_2;z)\mathcal{M}(k_3;z)}
        {\mathcal{M}(k_1;z)
          \mathcal{M}(k_2;z)}+\cyc\right]\nonumber\\
    &&+2F_2(\bfk_1,\bfk_2)P_0(k_1;z)P_0(k_2;z)+\cyc,
\label{eq:bis_pt}
\end{eqnarray}
where $\cyc$ denotes the cyclic permutations over the indices and
$P_0$, $B_0$, and $T_0$ are the power-, bi-, and tri-spectra of
$\deltam^{(1)}$.  In equation (\ref{eq:pow_pt}), the second term is
the first non-trivial correction in the presence of primordial
non-Gaussianity, and the function $B_0$ implies the primordial
bispectrum, which corresponds to the leading-order contribution in
equation (\ref{eq:bis_pt}).  Note that the contribution coming from
the primordial trispectrum $T_0$ is small for local type
non-Gaussianity with reasonable values of $\fnl$, and we drop this
term in computing $\Pm$ (see \cite{Taruya2008}).

On the other hand, the power spectrum of biased tracers has been
recently discussed in the literature
\cite{Dalal2008,Matarrese2008,Slosar2008,Afshordi2008,McDonald2008,Taruya2008},
based on several different formalisms including peak bias, halo bias
according to the peak-background split, and local bias. The resultant
expressions of the galaxy/halo power spectrum are basically the same,
and are summarized in the form
\begin{eqnarray}
\Pg(k;z) &=& b_1^2\left\{1\;+\;2\;\frac{\tilde{b}_2}{b_1}\;
\fnl\;\mathcal{M}^{-1}(k;z)\right\}^2\;P_0(k;z),
\label{eq:pow_bias}
\end{eqnarray}
where $b_1$ and $\tilde{b}_2$ are the bias parameters relating the
galaxy overdensity to the matter overdensity. In \ref{app:pow_bis}, we
present a derivation of (\ref{eq:pow_bias}) based on the local bias
formalism. The explicit expressions for the bias parameters $b_1$ and
$\tilde{b}_2$ can be obtained both from the peak bias and halo bias
formalisms, and their results are basically the same in the
high-peak/threshold limit. In \ref{app:bias}, we show that in the
high-peak limit, there is a clear relationship between the peak bias
and the local bias prescriptions, and the parameters $b_1$ and
$\tilde{b}_2$ are related to each other in terms of the critical
density of the spherical collapse model, $\deltac\approx1.686$, as
$\tilde{b}_2 =\deltac\;(b_1-1)$.

In equation (\ref{eq:pow_bias}), the factor in the braces manifestly
depends on the scale, which is the main source for ``scale-dependent
bias''. On large scales ($k\to0$), the function $\mathcal{M}$ is
roughly proportional to $k^2$, and it strongly affects the galaxy
power spectrum.  In this respect, the scale-dependent property will be
a clear indicator of primordial non-Gaussianity of the local type,
and it has been extensively tested against $N$-body
simulations. Several recent studies have suggested that some
modifications to this formula are required in order to model the
scale-dependent bias more accurately
\cite{Pillepich2008,Desjacques2009a,Grossi2009}.  For example,
\cite{Grossi2009} proposed a slight modification to the relation of
bias parameters, which reproduces results from $N$-body simulations very well:
\begin{eqnarray}
\tilde{b}_2 = \deltac\;q\;(b_1-1),
\label{eq:q_grossi}
\end{eqnarray}
with $q=0.75$, which comes from the ellipsoidal collapse model.

Now, we turn our focus to the bispectrum of biased tracers in the presence
of primordial non-Gaussianity.  According to the analytical study by
\cite{Jeong2009} (see also \cite{Sefusatti2009}), the effect of
local-type primordial non-Gaussianity is mainly imprinted on the
bispectrum of squeezed triangular configurations. The galaxy
bispectrum is then expressed as
\begin{eqnarray}
    \hspace{-0.5cm}\Bg(k,\alpha;z) &=& \Bg^{(0)}(k,\alpha;z) + \fnl\,\Bg^{(1)}(k,\alpha;z)
    + \fnl^2\,\Bg^{(2)}(k,\alpha;z),
\label{eq:Bsqueeze}
\end{eqnarray}
where each term of the right-hand side of this equation has the
following asymptotic form:
\begin{eqnarray}
\hspace{-0.5cm}\Bg^{(0)}(k,\alpha;z) &\simeq& b_1^2\,b_2\,P_{k\to0}^2(z)\,k^{2n_s}\,\alpha^0,
\label{eq:B0}\\
\hspace{-0.5cm}\Bg^{(1)}(k,\alpha;z) &\simeq& \left[4b_1^3+\left (\frac{26}{7} +
{\cal I}(k, \alpha; R) \right) b_1^2\tilde{b}_2\right]\,P_{k\to0}^2(z)
\,{\cal M}_{k\to0}^{-1}(z)\,k^{2n_s-2}\,\alpha^1,
\nonumber\\
\label{eq:B1} \\
\hspace{-0.5cm}\Bg^{(2)}(k,\alpha;z) &\simeq& 8\,b_1^2\,\tilde{b}_2\,P_{k\to0}^2(z)\,{\cal M}_{k\to0}^{-2}(z)\,k^{2n_s-4}\,\alpha^3.
\label{eq:B2}
\end{eqnarray}
Here, we focused on the isosceles triangles, and parametrized their
dependence as $k\equiv k_1= k_2 \equiv \alpha k_3$, and $n_{\rm s}$
denotes the scalar spectral index. See \ref{app:pow_bis} for a more
rigorous expression.  The function ${\cal I}(k, \alpha; R)$ weakly
depends on $k, \alpha$ and the smoothing scale, $R$, and it can be
approximated as $26/7 + {\cal I}(k, \alpha; R) \sim 34$ on large
scales. In the above, we take the limit of $k\to0$ and $\alpha\gg1$,
and $P_{k\to0}$, $\mathcal{M}_{k\to0}$ are defined through $P_0(k;z)
\to P_{k\to0}(k)k^{n_{\rm s}}$,
$\mathcal{M}(k;z)\to\mathcal{M}_{k\to0} (z)k^2$. See \ref{app:pow_bis}
and also \cite{Jeong2009} for more details.

Similar to the scale dependence of the galaxy/halo power spectrum in
equation (\ref{eq:pow_bias}), the amplitude of bispectrum is also
affected by the primordial non-Gaussianity in a scale-dependent
way. In particular, the term $\Bg^{(2)}$, which is of quadratic
order in $\fnl$, becomes dominant on large scales and exhibits a
behavior strongly dependent on $k$ and $\alpha$. Thus, we might 
conclude that this is the most important indicator of $\fnl$, capable of detecting the
primordial non-Gaussianity with upcoming galaxy surveys of large
volumes.

In what follows, we will examine this scale dependence in the
bispectrum of simulated dark matter haloes, with particular attention
on the squeezed configurations on large scales.

\section{$N$-body Simulations}
\label{sec:simulation}
\subsection{Setup}

We adopt the WMAP5 best-fit flat $\Lambda$CDM model \cite{WMAP5}:
$\Omega_{\rm m}=0.279$, $\Omega_\Lambda=0.721$, $\Omega_{\rm
  b}=0.046$, $h=0.701$, $\sigma_8=0.817$ and $n_s=0.96$, where
$\Omega_{\rm m}$, $\Omega_\Lambda$ and $\Omega_{\rm b}$ are the matter
density, cosmological constant and the baryon density normalized by
the critical density, $h$ is Hubble constant normalized by
$100\,$km$\,$s$^{-1}$Mpc$^{-1}$, $\sigma_8$ is the r.m.s. linear
density fluctuation smoothed by a top hat window function with radius
of $8h^{-1}$Mpc and $n_{\rm s}$ is the scalar spectral index.  We
calculate the linear matter transfer function using the {\tt CAMB}
code \cite{CAMB}. We have completed a total of $140$ realizations of 
matter clustering data: $20$ per each of the seven values for the local-type 
primordial non-Gaussianity parameter, $\fnl=0$, $\pm100$, $\pm300$ and $\pm1000$.
All the simulations were ran with the {\tt Gadget2} code
\cite{GADGET2}. We adopt $N=512^3$ particles in boxes of side $2000h^{-1}$Mpc, 
and set the softening length being $0.2h^{-1}$Mpc.
The parameters adopted in the simulations are the same
as in \cite{Taruya2009,Nishimichi2009}. We have tested the mass/force
resolution by changing the box size and found that the
results are well converged at large scales ($\sim1\%$ accuracy at $k\simlt0.3h$Mpc$^{-1}$)
for the matter power spectrum.

In setting the initial conditions, we first generate a random Gaussian field,
$\Phi_{\rm G}$, whose power spectrum is proportional to $k^{n_{\rm
    s}-4}$. We then apply the inverse Fourier transform, and add the
non-Gaussian contributions in real space according to equation
(\ref{eq:fnllocal}). Finally the real-space quantity is transformed
back to the Fourier space, and converted to the linear density
fluctuations by multiplying $\mathcal{M}(k;z)$ defined in equation
(\ref{eq:M}). We use second-order Lagrangian perturbation theory to
calculate the displacement field for $512^3$ particles placed on a
regular lattice \cite{Crocce2006} at $z=31$.

We store outputs at $z=2$, $1$ and $0.5$, and identify haloes for each
output using a FOF group finder with linking length of $0.2$ times the
mean separation. We select haloes in which the number of particles,
$N$, is equal to or larger than $10$, corresponding to the haloes with
masses $4.6\times10^{13}h^{-1}M_\odot$.  We also analyze haloes with
$N\geq20$ and $N\geq30$ to see the dependence on halo mass.

\begin{figure}[htbp]
\begin{center}
\includegraphics[height=12cm]{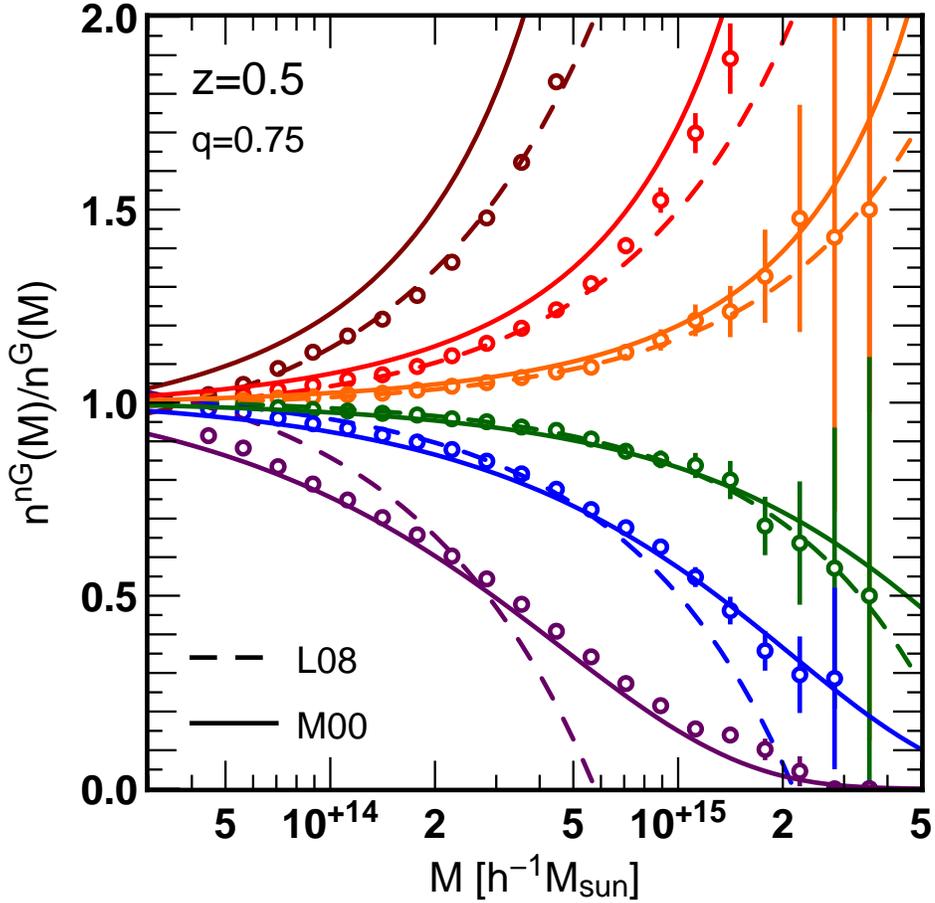}
\caption{ The ratio of the halo mass functions for non-Gaussian and
  Gaussian initial conditions at $z=0.5$.
  The symbols show the measurements from our simulations, while the
  lines are equations (\ref{eq:mf_l}) and (\ref{eq:mf_m}). The values of $\fnl$ are $1000$,
  $300$, $100$, $0$, $-100$, $-300$, and $-1000$ from top to bottom.}
\label{fig:massfunc}
\end{center}
\end{figure}

As a first check of the reliability of our simulations, we show in Fig.~\ref{fig:massfunc} 
the ratio of halo mass function with and without primordial non-Gaussianity, 
$R(M;\fnl)\equiv n_{\rm nG}(M;\fnl)/n_{\rm G}(M)$, at $z=0.5$, where different 
colors correspond to different values of 
$\fnl$: $1000$, $300$, $100$, $0$, $-100$, $-300$ and $-1000$ from top to bottom.
Both simulations and theoretical models 
suggest that the local type non-Gaussianity alters the mass function 
at the high-mass tail in the literature: a positive (negative) $\fnl$
enhances (suppresses) the tail. As shown in Fig.~\ref{fig:massfunc},
our simulations agree well with previously proposed analytical models.
The plotted lines show the model proposed by \cite{LoVerde2008} (solid):
\begin{eqnarray}
R(M;\fnl) = 1 + \frac{1}{6}\frac{\sigmar^2}{\delta_{\rm ec}}
\left[S_3
\left(\frac{\delta_{\rm ec}^4}{\sigmar^4}-2\frac{\delta_{\rm ec}^2}{\sigmar^2}-1\right)
+\frac{\rmd (\sigmar S_3)}{\rmd \ln\sigmar}\left(\frac{\delta_{\rm ec}^2}{\sigmar^2}-1\right)
\right],
\label{eq:mf_l}
\end{eqnarray}
based on the Edgeworth expansion to the probability
density function, and by \cite{Matarrese2000} (dashed):
\begin{eqnarray}
R(M;\fnl) = \exp\left(\frac{\delta_{\rm ec}^3}{6}\frac{S_3}{\sigmar^2}\right)
\left|\frac{1}{6}\frac{\delta_{\rm ec}}{\sqrt{1-\delta_{\rm ec}S_3/3}}
\frac{\rmd S_3}{\rmd \ln\sigmar}+\sqrt{1-\delta_{\rm ec}S_3/3}\right|,
\label{eq:mf_m}
\end{eqnarray}
obtained by the saddle-point approximation to the level excursion probability. 
In the above, $\sigmar\equiv\langle\deltam^2\rangle^{1/2}$ and 
$S_3\equiv\langle\deltam^3\rangle/\sigmar^4$.
These quantities are given as the function of mass $M$ 
through the relation $M=(4\pi/3)\overline{\rho}_{\rm
  m}\,R^3$, and linearly extrapolated to $z=0.5$. We also define 
  $\delta_{\rm ec} \equiv q\deltac$ again motivated by ellipsoidal collapse
model \cite{Grossi2009}. See also \cite{Desjacques2009a} for another model 
designed to fit to their $N$-body simulations. 

There are several claims on the systematics for the estimated
halo mass by FOF, and in fact our mass function fits better with 
\cite{Matarrese2000} when we correct that effect using the empirical 
formula of \cite{Warren2006}. We conclude here that our halo catalog
is accurate enough to investigate its clustering statistics.

\subsection{Measurements of the power spectrum and bispectrum}

Here, we briefly mention how to measure the power spectrum and
bispectrum in our simulations.

We assign particles (or haloes) to $1024^3$ grid points using the
Cloud-in-Cells (CIC) algorithm \cite{Hockney1981}.  We then Fourier
transform the density field, and divide each mode by the Fourier
transform of the CIC kernel to correct for the effect of assignment.
We made sure that the results are well converged at the scale of our
interest by changing the number of grid points.  We logarithmically
divide the measured power spectrum and bispectrum into wave number
bins starting from $k_{\rm min}=0.003h$Mpc$^{-1}$ and with $10$ bins
per decade. We select ``isosceles'' triangles, whose two longer sides,
$k_1$ and $k_2$, fall into the same bin for the bispectrum analysis.
In this sense our isosceles triangles are not strictly isosceles, but
we adopt this convention to reduce the statistical errors on the
measured bispectrum caused by small number of triangles in
$k$-space. In plotting results, we assign each data point to the
logarithmic-central value of wave number in that bin.

\section{Results of Power Spectrum}
\label{sec:pow_results}

In this section, we present the power spectrum measured from $N$-body
simulations We first compare the measured matter power spectrum with
the perturbation theory prediction. We then present the halo power
spectrum, and compare it with the analytic models proposed in the
literature. These are important sanity checks to justify the
results of our simulations in subsequent sections. 

\subsection{matter power spectrum}

\begin{figure}[htbp]
\begin{center}
\includegraphics[width=12cm]{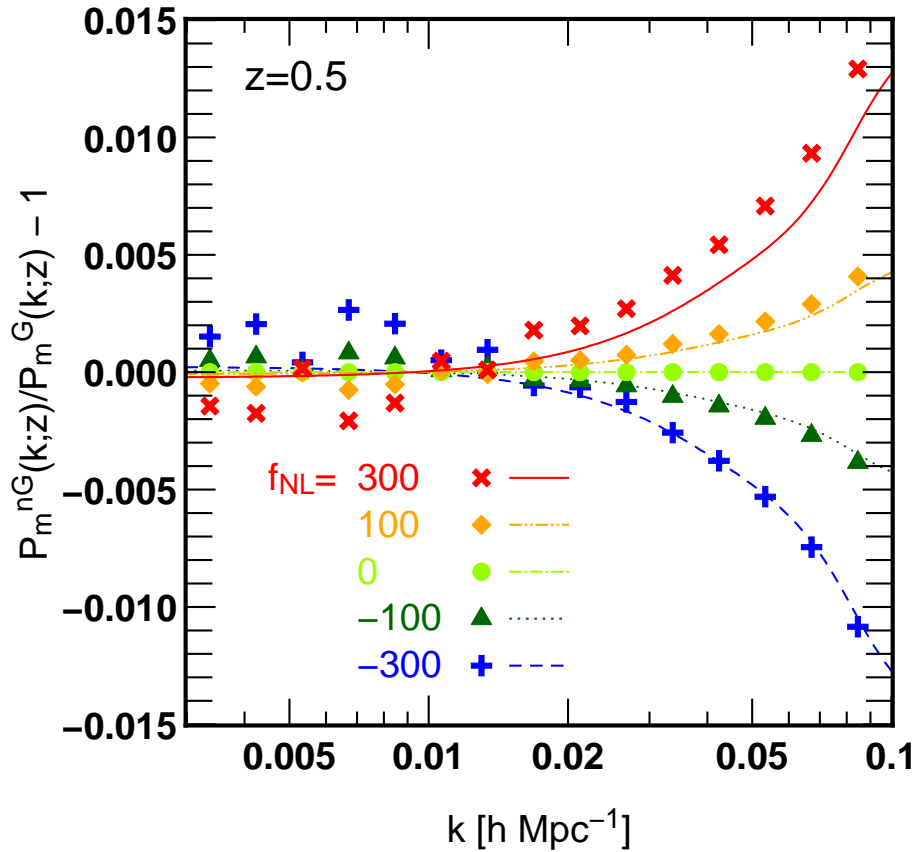}
\caption{Fractional differences of the matter power spectra starting
  from non-Gaussian and Gaussian initial conditions. Symbols show the
  results of $N$-body simulations, while lines are perturbation theory
  predictions of equation (\ref{eq:pow_pt}) ($\fnl=300,100,0,-100,
  -300$ from top to bottom at $k\simgt0.02h$Mpc$^{-1}$).}
\label{fig:dmpow}
\end{center}
\end{figure}

We first examine the matter power spectrum, in cases with non-zero
$\fnl$.  Fig. \ref{fig:dmpow} shows the fractional difference of the
matter power spectra between Gaussian and non-Gaussian initial
conditions measured at $z=0.5$. The symbols represent $\fnl=300, 100,
0, -100$ and $-300$ from top to bottom at
$k\simgt0.02h$Mpc$^{-1}$. Since we use the same set of random seeds
for the seven $\fnl$ parameters, the cosmic variance is effectively
canceled out by taking the ratios, $P_{\rm m}^{\rm nG}(k;z)/P_{\rm
  m}^{\rm G}(k;z)$. Overall, the deviations from Gaussian results
themselves are very small (less than $1\%$ when $|\fnl|=100$ in the
plotted range). 

In Fig. \ref{fig:dmpow}, we also plot the predictions based on
perturbation theory of \cite{Taruya2008}, depicted as continuous
lines. Note that recently, Ref.~\cite{Bartolo2010} developed another
analytical model based on the Time-RG approach, which would be more accurate
in the weakly nonlinear regime.  However, the standard perturbation 
theory prediction of \cite{Taruya2008} is 
accurate enough at the scale of our interest (i.e., $k\simlt0.05h$Mpc$^{-1}$),
as was shown by comparisons with $N$-body simulations in Refs.~\cite{Pillepich2008,Desjacques2009a}.

Fig.~\ref{fig:dmpow} shows that the results of our $N$-body simulations are in reasonably 
good agreement with the model of \cite{Taruya2008}, except for the case of a large
non-Gaussianity with $\fnl=300$.  The discrepancy between $N$-body and
analytic results seen in the $\fnl=300$ case might be partially
ascribed to the term coming from the primordial trispectrum in
equation (\ref{eq:pow_pt}), which is neglected in the perturbation
theory calculations. Nevertheless, the discrepancy remains at the 
sub-percent level, and thus does not seriously affect the later
analysis of the matter/halo bispectrum.

\subsection{halo power spectrum}

\begin{figure}[htbp]
\begin{center}
\includegraphics[width=12cm]{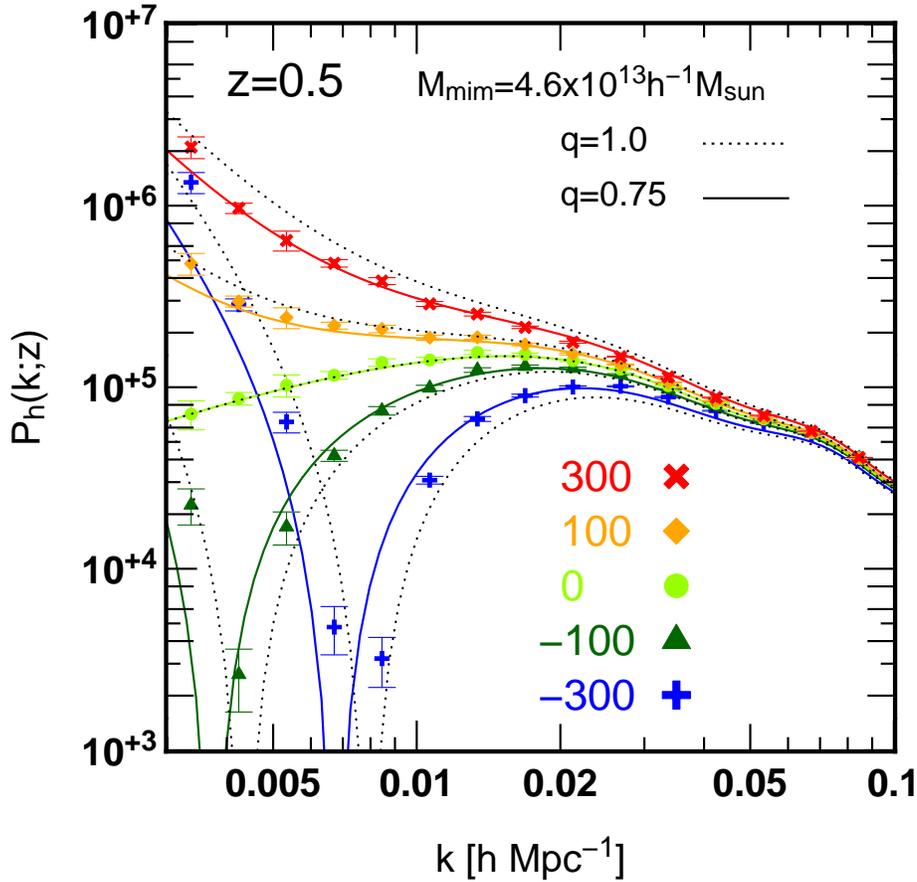}
\caption{The power spectrum of haloes more massive than
  $4.6\times10^{13}h^{-1}M_\odot$ at $z=0.5$. Symbols are results of
  $N$-body simulations, while lines are theoretical prediction of
  equation (\ref{eq:pow_bias}) with (\ref{eq:q_grossi}) where we adopt
  $q=1.0$ for dotted lines and $q=0.75$ for solid lines
  ($\fnl=300,100,0,-100, -300$ from top to bottom).}
\label{fig:pg_all}
\end{center}
\end{figure}

We next consider the halo power spectrum for various values of $\fnl$,
shown in Fig. \ref{fig:pg_all}. We set the minimum mass of haloes to
$4.6\times10^{13}h^{-1}M_\odot$, which corresponds to $10$ $N$-body
particles. We also show the analytical prediction of equation
(\ref{eq:pow_bias}) with the bias parameter $b_1$ fitted to reproduce
the results of Gaussian simulations, adopting the relation between
bias parameters in equation (\ref{eq:q_grossi}). We plot the model
with $q=1$ and $q=0.75$ by dotted and solid lines, which corresponds
to the original peak bias prediction and the fit by \cite{Grossi2009},
respectively.

Overall, the scale dependence of the halo power spectrum discussed in
the literature can be clearly seen in our simulations with very small
statistical errors, owing to the large total volumes. The agreement
between $N$-body simulations and the analytic models becomes better
when we choose $q=0.75$, consistent with \cite{Grossi2009}.

Note, however, that the choice of $q=0.75$ does not necessarily imply
the best-fit results: $q=0.85$ gives a better fit to this particular
case, and the best-fit value of $q$ changes with redshift and minimum
halo mass. This might indicate that there exists some systematic
effects on the halo clustering properties in our simulations. One
possibility is the difference in the halo finding algorithms: while we
adopt the FOF finder, \cite{Grossi2009} use a SO finder. See also
\cite{Pillepich2008}, where they use a FOF finder and proposed a fit
corresponding to $q=0.8$.

We may further improve the agreement between $N$-body simulations and
theoretical predictions by including some corrections to the theory.
Ref. \cite{Desjacques2009a} showed that the inclusion of two
corrections coming from the changes in halo mass function and the
matter power spectrum actually improves their results. These
systematics will definitely be important for the application to the
upcoming surveys, however, we do not pursue this issue in the present
paper, since our primary focus is on the halo bispectrum.

\section{Results of Bispectrum}
\label{sec:bis_results}

In this section, we present the bispectrum measured from $N$-body
simulations.  Throughout the analysis, we consider the isosceles
triangles for the configuration of bispectrum, which are characterized
by the two parameters $k$ and $\alpha$, defined by $k\equiv k_1 = k_2
\equiv \alpha k_3$. We pay special attention to the squeezed
triangles, $\alpha\gg1$. We first present the results of the matter
bispectrum (Sec.~\ref{subsec:matter_bispec}), and then discuss how the
halo bispectrum differs from the matter bispectrum
(Sec.~\ref{subsec:halo_bispec}). While we mainly analyze the default
halo catalog with minimum mass $M_{\rm
  min}=4.6\times10^{13}h^{-1}M_\odot$ and output redshift $z=0.5$, we
briefly discuss how the results are changed when we vary the minimum
halo mass and redshift (Sec.~\ref{subsubse:mass_redshift}).

\subsection{matter bispectrum}
\label{subsec:matter_bispec}

\begin{figure}[htbp]
\begin{center}
\includegraphics[width=12cm]{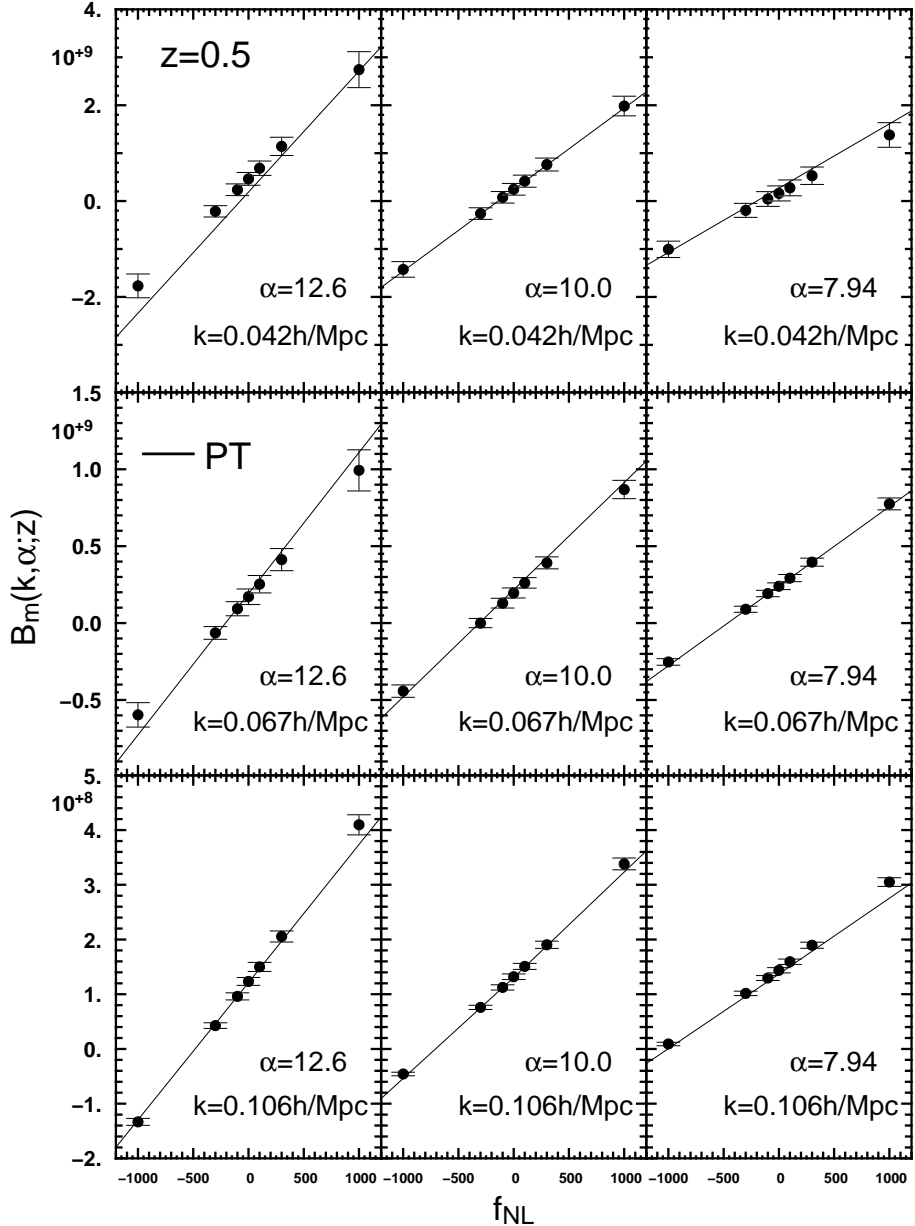}
\caption{The matter bispectrum. Each panel shows the results for an
  isosceles configuration specified by $\alpha\equiv k_1/k_3$ and
  $k\equiv k_1=k_2$. Symbols are measurements from $N$-body
  simulations (the average and the standard error among different
  realizations) and solid lines are the perturbation theory
  predictions of equation (\ref{eq:bis_pt}).}
\label{fig:bm}
\end{center}
\end{figure}

Let us present the results of the measured matter bispectrum.  In
Fig.~\ref{fig:bm}, the symbols in each panel show the amplitude of the 
bispectrum measured from simulations for various $\fnl$ at a fixed
triangle specified by $k$ and $\alpha$ indicated in the panel.  We
also show the perturbation theory prediction of equation
(\ref{eq:bis_pt}) by solid lines. Note that the value of $\alpha$
increases from right to left, while $k$ increases from top to bottom.

Although we have very large total volume, the statistical uncertainty
due to finiteness of the simulated volume still affects the measurements.
We thus take account of this effect in the perturbation theory predictions:
we compute the matter bispectrum using the second-order perturbation theory 
starting from linear density field realized in finite-volume boxes which were 
used to generate the initial conditions of the simulations, and take average over 
realizations. Namely, we compute
\begin{eqnarray}
&&\hspace{-2cm}{\rm Re}\Bigl[\deltam^{(1)}(\bfk_1)\deltam^{(1)}(\bfk_2)\deltam^{(1)}(\bfk_3)\nonumber\\
&&\hspace{-2cm}+\deltam^{(2)}(\bfk_1)\deltam^{(1)}(\bfk_2)\deltam^{(1)}(\bfk_3)+\deltam^{(1)}(\bfk_1)\deltam^{(2)}(\bfk_2)\deltam^{(1)}(\bfk_3)+\deltam^{(1)}(\bfk_1)\deltam^{(1)}(\bfk_2)\deltam^{(2)}(\bfk_3)\Bigr],
\end{eqnarray}
and take the average over the realizations and triangles in the bin 
for the perturbation theory prediction.
As a result, the analytical predictions are in good agreements with
measurements from simulations.

The matter bispectrum from both simulations and perturbation theory
clearly exhibits a linear dependence on $\fnl$ for all the configurations plotted in
Fig.~\ref{fig:bm}. Based on this results, we will discuss how the
dependence on $\fnl$ is modified for the halo bispectrum.

\subsection{halo bispectrum}
\label{subsec:halo_bispec}

We are now in a position to show the halo bispectrum.  Since this is the
first numerical study on the halo bispectrum in the presence of local
type non-Gaussianity, it is important to understand the $N$-body
results in a model-independent manner.  In this subsection, we first
show the $\fnl$ dependence of the halo bispectrum. We then consider
the scale dependence and compare the simulation results with the
theoretical prediction of \cite{Jeong2009}. The dependence on the
minimum halo mass and redshift is also investigated in section
\ref{subsubse:mass_redshift}.

\subsubsection{$\fnl$ dependence}

In order to quantitatively study the $\fnl$ dependence of the halo
bispectrum, we use all the halo catalogs with various values of
$\fnl$, and fit the measured bispectrum to the polynomial form:
\begin{eqnarray}
\Bh(k,\alpha;z|\fnl) = \sum_{i=0}^{4}\fnl^i\Bh^{(i)}(k,\alpha;z),
\label{eq:bh_fit}
\end{eqnarray}
using the standard $\chi^2$ fitting with the variance of the 
data points measured from $N$-body simulations.
For specific configurations of $(k,\alpha)$, we determine the
parameters $\Bh^{(i)}$ using the halo catalogs with different values
of $\fnl$. We confirmed that the results are almost unchanged when we
add higher-order polynomials with $i\geq5$.

\begin{figure}[htbp]
\begin{center}
\includegraphics[width=12cm]{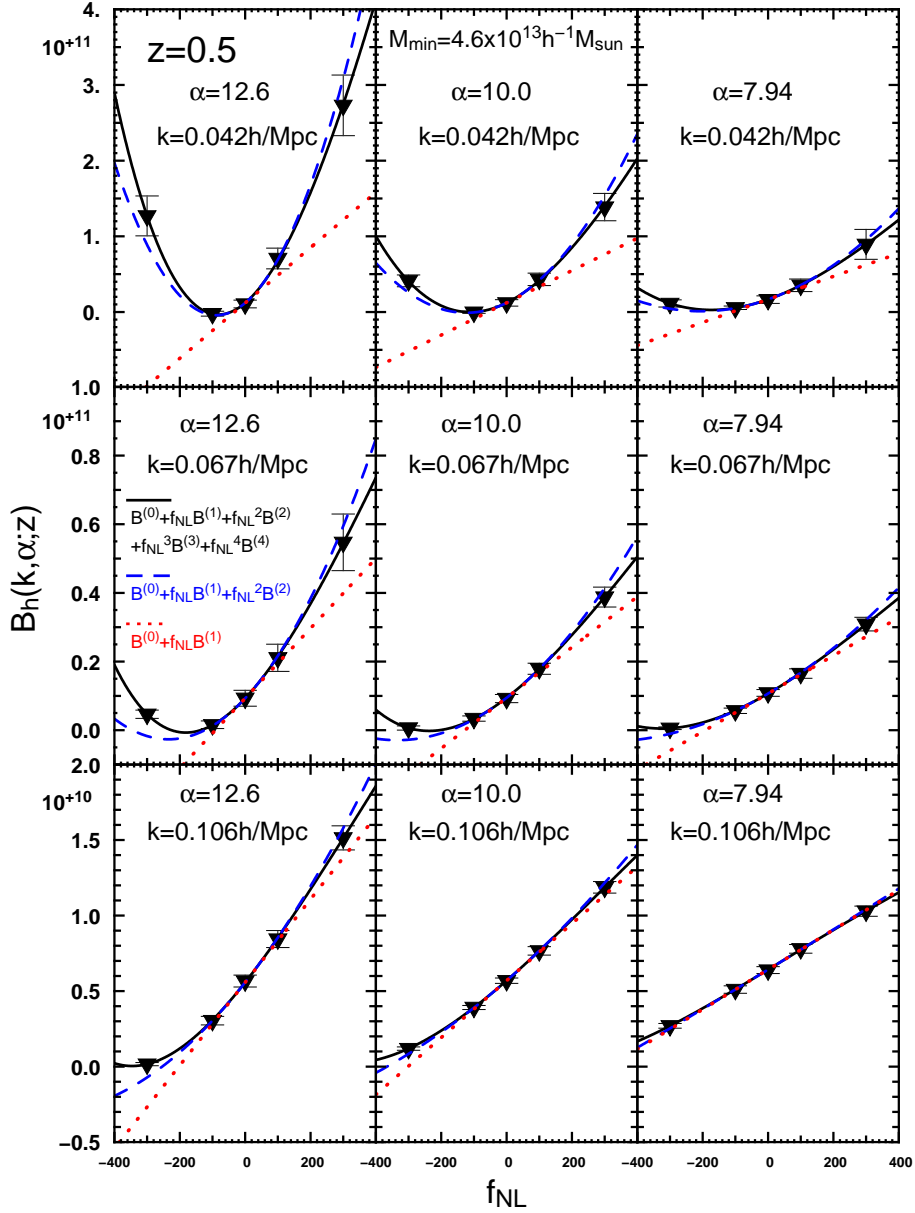}
\caption{The halo bispectrum for some triangular configurations. Each
  panel shows the result for an isosceles configuration specified by
  $\alpha\equiv k_1/k_3$ and $k\equiv k_1=k_2$. Error bars are
  measurements from our simulations (the average and the standard
  error among different realizations) and solid lines are their $4$-th
  order polynomial fits, while we keep the terms up to second and
  linear order for dashed and dotted lines. We use the outputs at
  $z=0.5$ and consider the haloes more massive than
  $4.6\times10^{13}h^{-1}M_\odot$.}
\label{fig:bh}
\end{center}
\end{figure}

In Fig.~\ref{fig:bh}, we show the measured bispectrum as function of
$\fnl$ for the same set of triangular configurations as plotted in
Fig.~\ref{fig:bm}.  Note again that the value of $\alpha$ increases
from right to left panels, while $k$ increases from top to bottom
panels.  The fitted results of Eq.~(\ref{eq:bh_fit}) truncating at the
first order ($i=0,1$), second order ($i=0\sim2$) and fourth order
($i=0\sim4$) are shown respectively as dotted, dashed and solid lines.
Although we do not show the points at $\fnl=\pm1000$ in order to focus
on more realistic values of $\fnl$, we take account of these results
when we fit the $N$-body data to Eq.~(\ref{eq:bh_fit}).

The second-order term ($\Bh^{(2)}$) becomes more significant in moving
from bottom to top and from right to left, and in the end, the
top-left panel ($k=0.042h$Mpc$^{-1}$, $\alpha=12.6$) shows strong
evidence of $\Bh^{(2)}$. Higher order terms ($\Bh^{(3)}$ and
$\Bh^{(4)}$) seem to have almost no effect on the total bispectrum
when $|\fnl|\simlt100$, although they may play some roles at
$\fnl=\pm300$ (and also $\pm1000$).  This second order term,
$\Bh^{(2)}$, is not seen in the matter bispectrum (Fig. \ref{fig:bm})
and we for the first time confirme that this really exists in the
halo clustering in $N$-body simulations.

\subsubsection{shape and scale dependence}

\begin{figure}[htbp]
\begin{center}
\includegraphics[width=7cm]{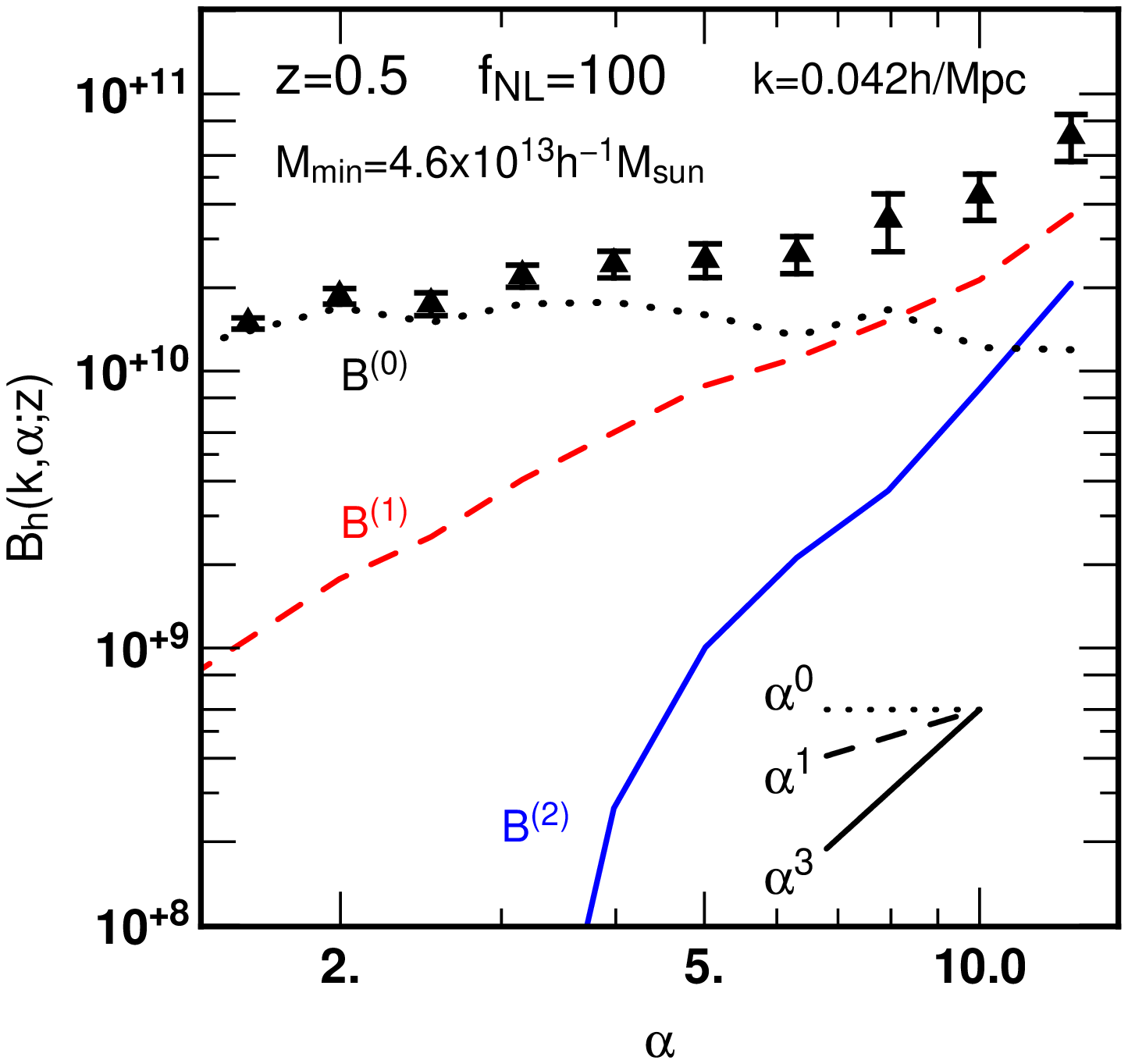}
\includegraphics[width=7cm]{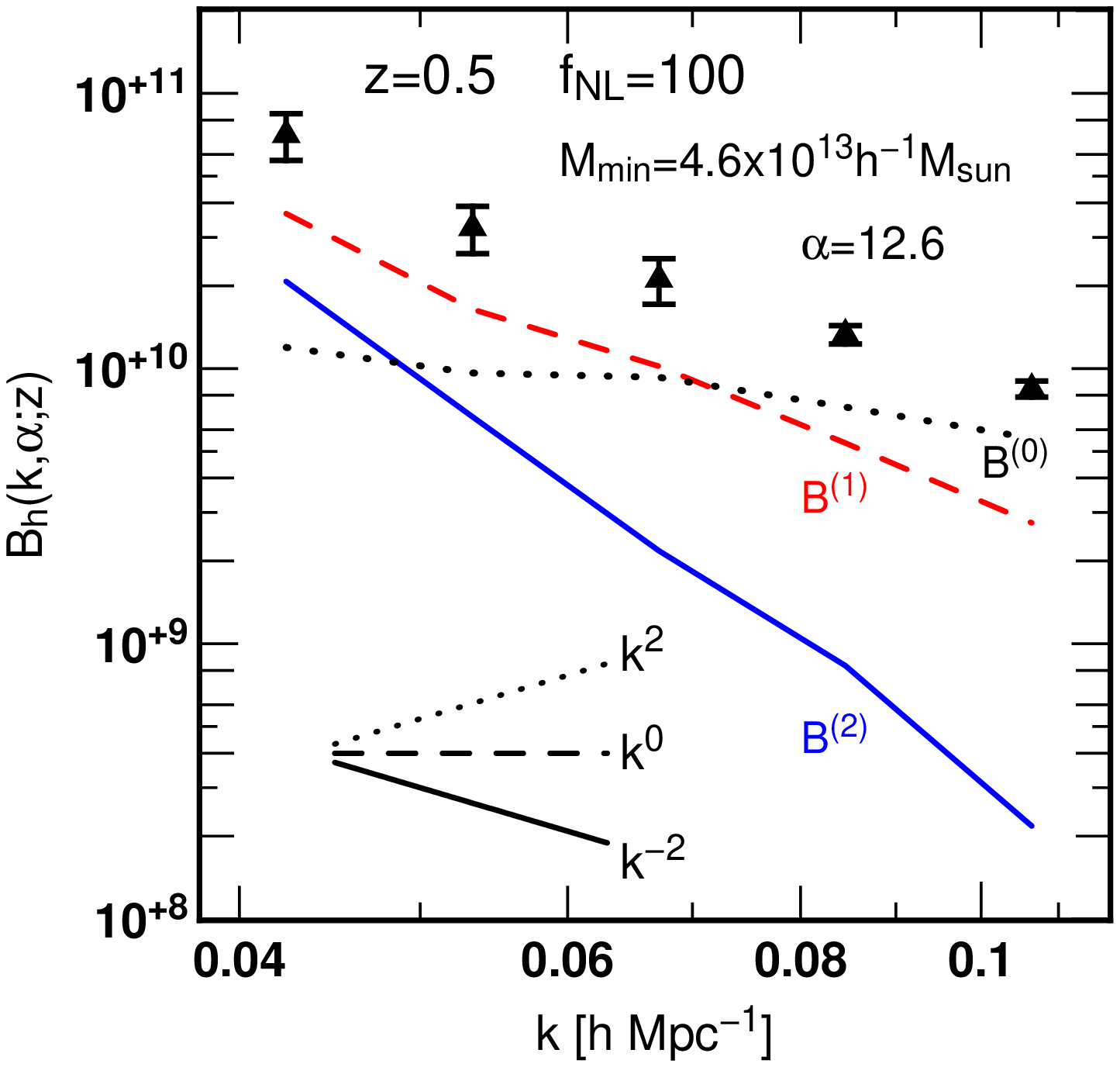}
\caption{Shape and scale dependence of the halo bispectrum. {\it
    left}: $\alpha$ dependence when $k$ is fixed, {\it right}: $k$
  dependence when $\alpha$ is fixed.  We plot the results at $z=0.5$
  for haloes more massive than $4.6\times10^{13}h^{-1}M_\odot$ when
  $f_{\rm NL}=100$.  Error bars are measurements from our simulations,
  while dotted, dashed and solid lines are terms which scale as
  $f_{\rm NL}^0$, $f_{\rm NL}^1$, and $f_{\rm NL}^2$, respectively.
  Short straight lines are corresponding analytical prediction in the
  squeezed limit.}
\label{fig:dependence}
\end{center}
\end{figure}

We next investigate the scale dependence of
$\Bh^{(i)}(k,\alpha;z)$. We show them in Fig.~\ref{fig:dependence} for
$\fnl=100$. The left panel shows the $\alpha$ dependence when the wave
number $k$ is fixed to $0.042h$Mpc$^{-1}$, while the right panel
displays the $k$ dependence when $\alpha=12.6$. The analytic
prediction based on local bias (see \S \ref{sec:theory} and
\ref{app:pow_bis}) predicts $\Bh^{(0)}\propto k^2\alpha^0$,
$\Bh^{(1)}\propto k^0\alpha^1$ and $\Bh^{(2)}\propto k^{-2}\alpha^3$
in the squeezed limit at large scales ($k\to0$, $\alpha\gg1$), and we
show these asymptotic scalings by short straight lines (normalizations
are arbitrary).

The $\alpha$ dependence in the left panel is quite consistent with the
theoretical predictions, and the results strongly indicate that the
theoretical model captures the nature of the shape dependence.  On the
other hand, the $k$ dependence measured from $N$-body simulations
seems different from that predicted by the theoretical model. This
implies that the wave numbers shown in the figure are not sufficiently
small, and the approximation used in deriving the theoretical
predictions is not valid.  We expect that the asymptotic scaling
appears only at the scales larger than the turn over of the power
spectrum (i.e., $T(k)\simeq1$).  In fact, the value of the matter
transfer function at $k=0.042h$Mpc$^{-1}$, corresponding to the
wavenumber at the left-most bin in the panel, is $T(k)=0.293$, and
thus the approximation of
$\mathcal{M}(k)\simeq\mathcal{M}_{k\to0}\;k^2$ used in equation
(\ref{eq:Bsqueeze}) cannot be applied. One can find a similar feature in Fig.~7
of Ref.~\cite{Sefusatti2009} where the authors compute the galaxy bispectrum
using one-loop perturbation theory adopting the local bias model. 
Although it illustrates the galaxy bispectrum for
equilateral triangles, the asymptotic power law feature appears only at very
large scale (i.e., $k\simlt0.03h$Mpc$^{-1}$).

Nevertheless, both simulations and theory suggest that 
$\Bh^{(2)} > \Bh^{(1)} > \Bh^{(0)}$ at the limit of small $k$. The $\Bh^{(2)}$ term will play an
important role in constraining $\fnl$ from future surveys where we can
investigate such large scales.

\subsubsection{dependence on halo mass and redshift}
\label{subsubse:mass_redshift}

\begin{figure}[htbp]
\begin{center}
\includegraphics[width=7cm]{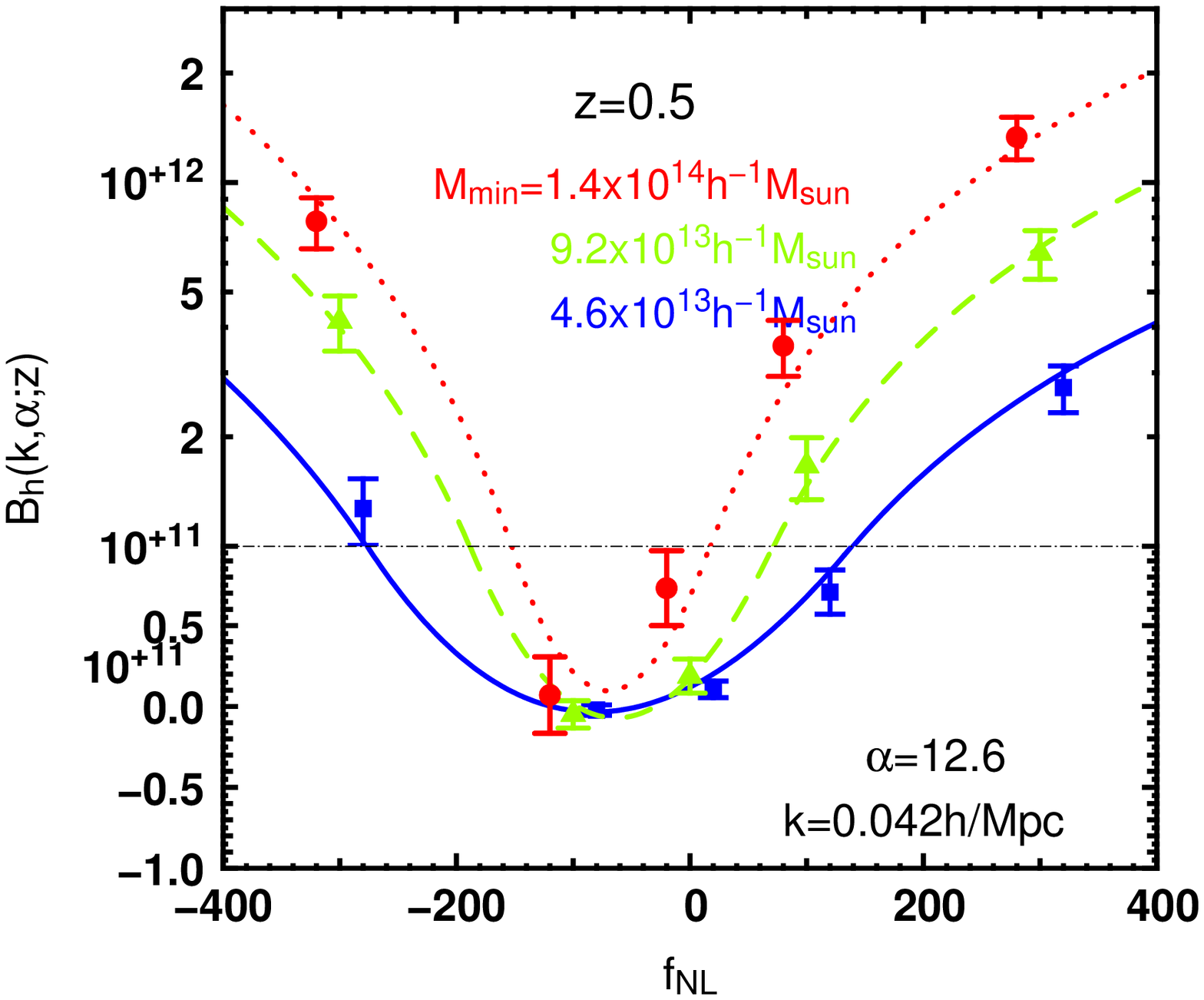}
\includegraphics[width=7cm]{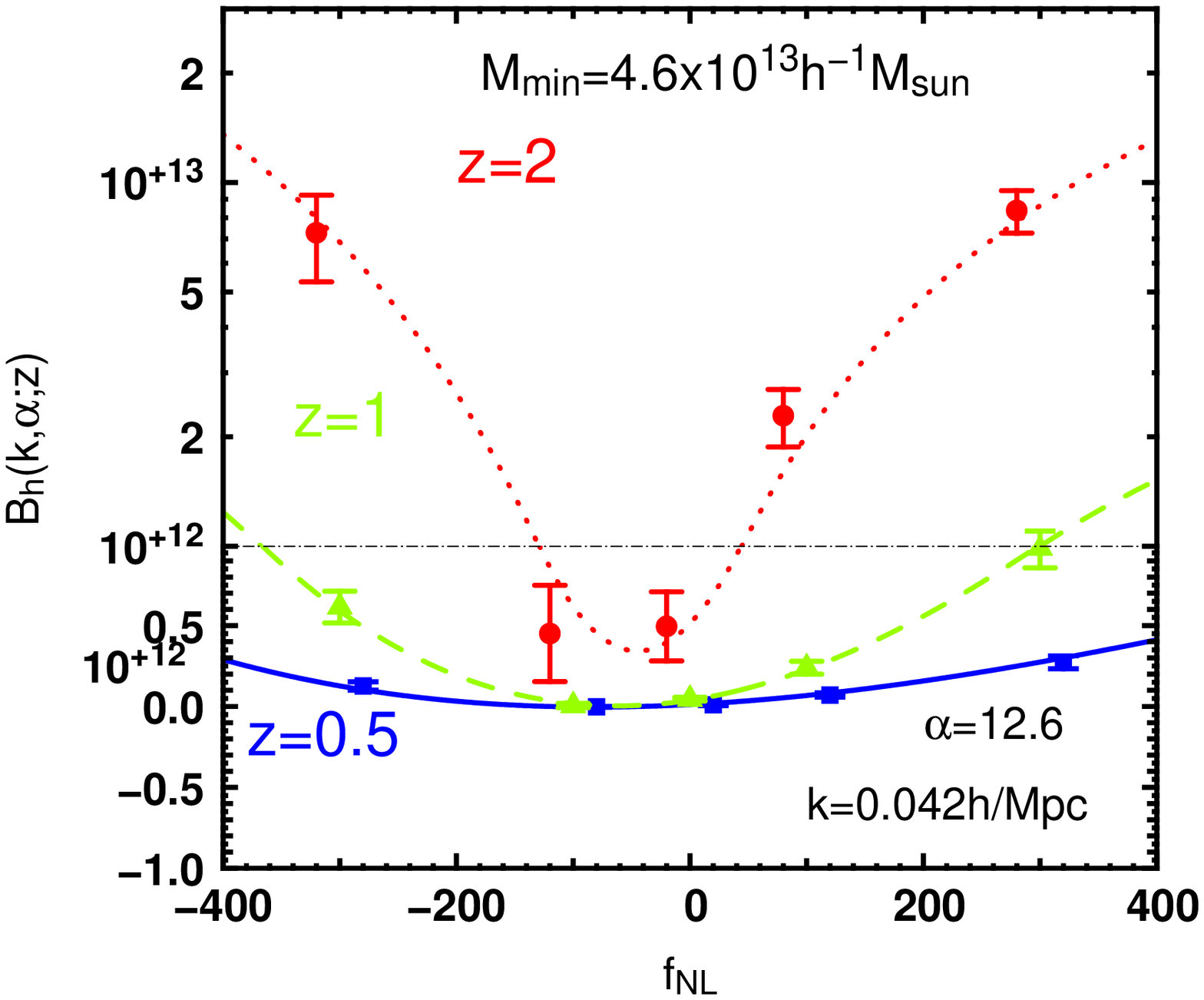}
\caption{{\it left}: Mass dependence of the halo bispectrum at
  $z=0.5$. Symbols and lines are similar to Fig. \ref{fig:bh}, but we
  fix the triangular configuration to be the same as the top-left
  panel, while changing the minimum halo mass: $4.6\times10^{13}$,
  $9.2\times10^{13}$ and $1.4\times10^{14}h^{-1}M_\odot$ for squares,
  triangles, and circles, respectively.  Note that although we do not
  show the results for $f_{\rm NL}=\pm1000$, we take them into account
  for the polynomial fitting. Note also that the vertical axis is
  logarithmic above the dot-dashed horizontal line, while it is linear
  below it.  {\it right}: same as left panel but the redshift
  dependence for a fixed minimum halo mass
  ($4.6\times10^{13}h^{-1}M_\odot$). Squares, triangles, and circles
  correspond to $z=0.5$, $1$, and $2$, respectively.}
\label{fig:bdepend_squeezed}
\end{center}
\end{figure}

So far, we have concentrated on the haloes with $M_{\rm
  halo}\geq4.6\times10^{13}h^{-1}M_\odot$ at $z=0.5$.  In this
subsection, we extend our analysis to the haloes with higher mass
thresholds and at different redshifts to see the dependence of the
halo bispectrum on these quantities.  For this purpose, we
specifically consider the squeezed triangle with $k=0.042h$Mpc$^{-1}$
and $\alpha=12.6$, corresponding to the configuration shown in the
top-left panel of Fig.~\ref{fig:bh}, and plot in
Fig.~\ref{fig:bdepend_squeezed} the amplitude of the bispectrum against
$\fnl$ for different mass thresholds (left) and redshifts (right).

In the left hand panel, each symbol and a line respectively correspond to the
measurements and the polynomial fit based on equation
(\ref{eq:bh_fit}) for a fixed minimum mass of haloes given by $M_{\rm
  min}=4.6\times10^{13}h^{-1}M_\odot$ (square/solid),
$9.2\times10^{13}h^{-1}M_\odot$ (triangle/dashed) and
$1.4\times10^{14}h^{-1}M_\odot$ (circle/dotted). Similar to
Fig.~\ref{fig:bh}, we can see a clear quadratic dependence on $\fnl$,
but the role of the quadratic term $B_{\rm h}^{(2)}$ seems more
significant for haloes with larger minimum masses.

In the right hand panel, each symbol and line respectively show the
measurements and a fit at the specific redshifts $z=0.5$
(square/solid), $1$ (triangle/dashed) and $2$ (circle/dotted).  Here,
we fix the minimum halo mass to $M_{\rm
  min}=4.6\times10^{13}h^{-1}M_\odot$.  Again, one can see the
quadratic dependence on $\fnl$, which become more important for higher
redshifts.

Although we do not try to find the best halo catalog or optimal
weighting scheme to detect the signal of $\fnl$ here, the balance
between having denser samplings and getting higher signals for $\fnl$
by selecting massive haloes is clearly very important. We will
investigate these issues elsewhere.

\section{Prospects for future survey}
\label{sec:discussion}

In this section, we discuss future prospects to detect $\fnl$ through
the measurements of the bispectrum. We especially pay attention to the
importance of the higher order term, $\Bg^{(2)}$ in equation
(\ref{eq:bg_simple}), which is defined in analogous to $\Bh^{(2)}$ in
equation (\ref{eq:bh_fit}) that scales as $\fnl^2$.

We consider three representative surveys: (i) idealistic survey with a
huge volume and a deep sampling ($V = 100h^{-3}$Gpc$^3$, $n_{\rm
  g}=1\times10^{-3}h^3$Mpc$^{-3}$, $z=1$), (ii) realistic
survey with a large volume accessible in near future ($V =
10h^{-3}$Gpc$^3$, $n_{\rm g}=5\times10^{-4}h^3$Mpc$^{-3}$,
$z=1$), and (iii) deep survey ($V = 3h^{-3}$Gpc$^{3}$, $n_{\rm
  g}=3\times10^{-4}h^3$Mpc$^{-3}$, $z=2$). 
Parameters of
these three surveys roughly correspond to EUCLID \cite{EUCLID}, SuMIRe
\cite{SuMIRe} and HETDEX \cite{HETDEX}, respectively, except for the
slightly smaller value of redshift in the deep survey.  Although the
mass resolution of the current simulations are not sufficient to
reproduce the same number density of galaxies in those surveys, it is
worth giving a rough estimate of the detectability.

Under the assumption of the one-to-one correspondence between haloes
and galaxies, we first estimate the minimum halo mass $M_{\rm min}$
that reproduces the mean galaxy number density, $n_{\rm g}$, for each
survey. We use the mass function of \cite{Crocce2009} to derive the
minimum value, $M_{\rm min}$. We then compute the linear bias
parameter, $b_1$, using the Sheth \& Tormen fit \cite{Sheth1999}.
The resultant minimum masses and bias parameters are summarized
in Tab. \ref{tab:survey}.
\begin{table}
\begin{center}
\caption{The survey parameters used for the forecasts.
We assume the one-to-one correspondence between haloes and galaxies
to compute the minimum halo mass and the bias parameter from the number
density.\\
\label{tab:survey}}
\begin{tabular}{c||c|c|c|c|c}
survey & $z$ & $V\,[h^{-3}$Gpc$^3]$ & $n_{\rm g}\,[h^3$Mpc$^{-3}]$ & $M_{\rm min}\,[h^{-1}M_\odot]$ & $b_1$\\
\hline
IDEAL & $1.0$ & $100$ & $1\times10^{-3}$ & 
$2.8\times10^{12}$ & $1.9$\\
\hline
REALISTIC & $1.0$ & $10$ & $5\times10^{-4}$ & 
$5.0\times10^{12}$ & $2.2$\\
\hline
DEEP & $2.0$ & $3$ & $3\times10^{-4}$ & 
$3.4\times10^{12}$ & $3.3$
\end{tabular}
\end{center}
\end{table}

Based on our numerical experiments, we focus on the isosceles
triangles with $k=0.042h$Mpc$^{-1}$ (see the left panel of
Fig.~\ref{fig:dependence}), and model the galaxy bispectrum similar to
the halo bispectrum (\ref{eq:bh_fit}) as
\begin{eqnarray}
    \hspace{-2cm} {\Bg}(\alpha;z,M_{\rm min}) &=&
    B^{(0)}(z,M_{\rm min})\Bigl[
    1+\fnl\alpha C^{(1)}(z,M_{\rm min})+\fnl^2\alpha^3 C^{(2)}(z,M_{\rm min})
    \Bigr].
    \label{eq:bg_simple}
\end{eqnarray}
We estimate the coefficients, $B^{(0)}$, $C^{(1)}$ and $C^{(2)}$ from
the three halo catalogs using the fitting procedure described in the previous section.  We then scale
them to the minimum halo masses for the three surveys as follows: for
the coefficient $B^{(0)}$, we scale as $\propto b_1^4$.  This is based
on the fact that the term in equation~(\ref{eq:B0}), which scales as
$b_1^2b_2$, is the dominant contribution at large scales, and
$b_2\propto b_1^2$ at the high-peak limit (see \ref{app:bias}). On the
other hand, we assume that $C^{(1)}$ and $C^{(2)}$ do not sensitively depend
on $M_{\rm min}$ and $z$, and treat them as constants.

\begin{figure}[htbp]
\begin{center}
\includegraphics[height=6.5cm]{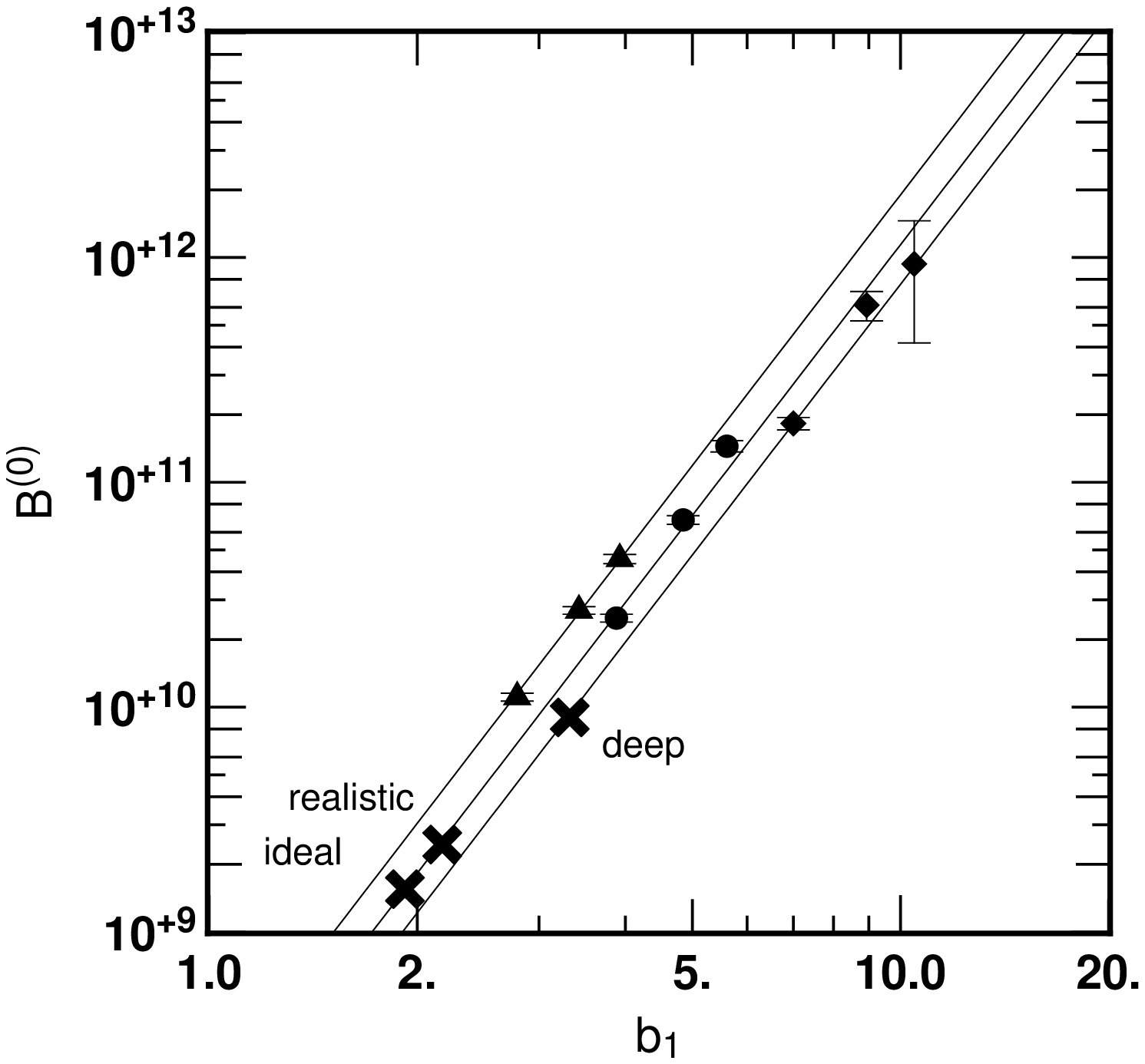}
\includegraphics[height=6.5cm]{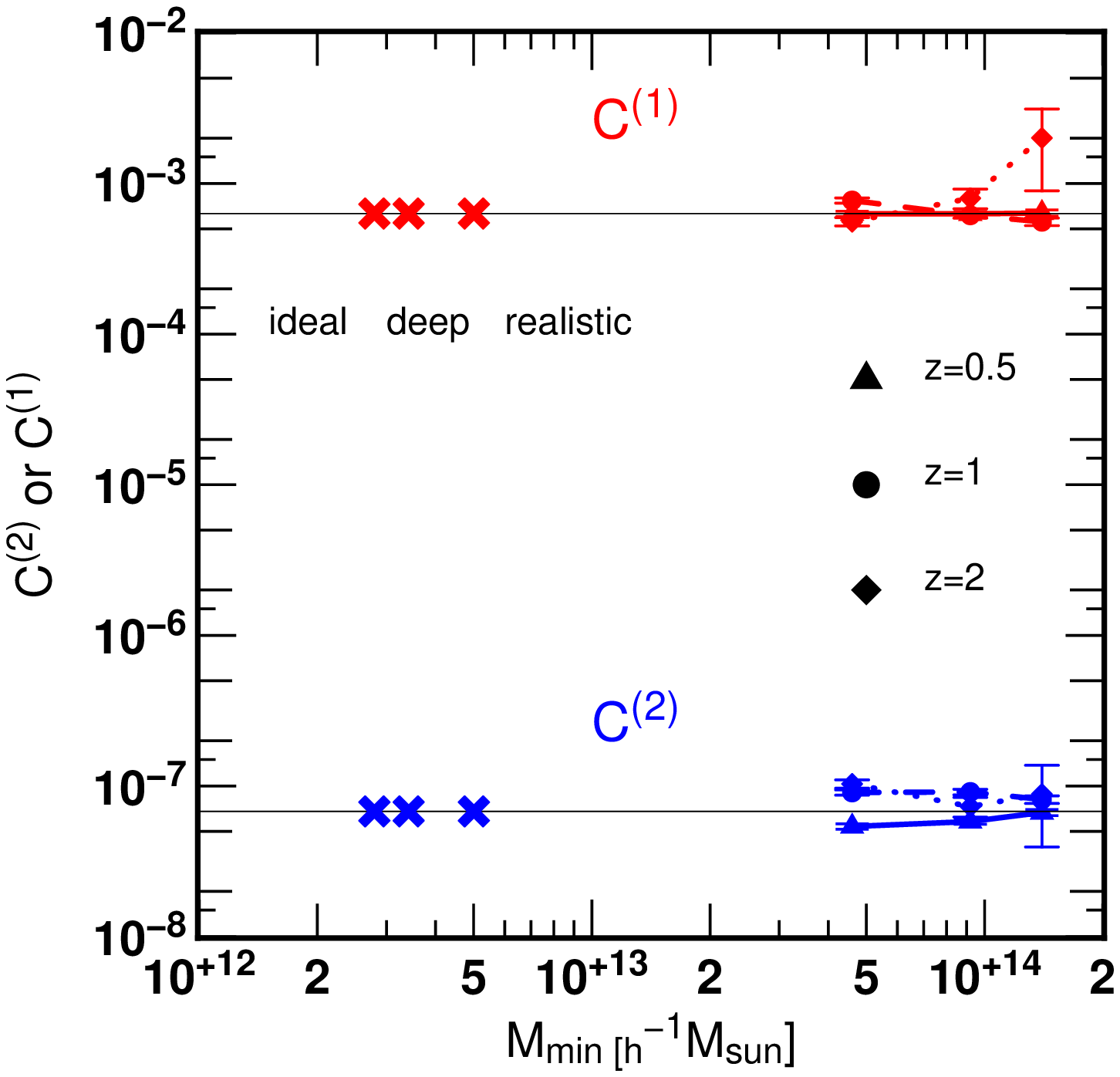}
\caption{Scalings of the coefficients in equation~(\ref{eq:bg_simple}).
  We assume that $B^{(0)}\propto b_1^4$, while $C^{(1)}$ and
  $C^{(2)}$ are treated as constants. We extrapolate these coefficients
  to the three surveys
  depicted by crosses. The triangles, circles and
  diamonds correspond to the measurements from $N$-body simulations at
  $z=0.5$, $1$ and $2$, respectively. 
  {\it left}: $B^{(0)}$ as a function of the linear bias. {\it right}: $C^{(1)}$
  and $C^{(2)}$ as a function of the minimum halo mass.
  }
\label{fig:bis_scaling}
\end{center}
\end{figure}

The left hand panel of Fig.~\ref{fig:bis_scaling} illustrates the scaling of $B^{(0)}$ 
measured from the $N$-body simulations. The triangles, circles and diamonds 
respectively correspond to the measurements from $N$-body simulations at 
$z=0.5$, $1$ and $2$, respectively. We also show the $\propto b_1^4$ scaling by 
three solid lines. The scaling seems to be a reasonable fit to the simulations.
In the right hand panel, we plot $C^{(1)}$ and $C^{(2)}$ as a function of the minimum halo
mass at the three redshifts ({\it upper}: $C^{(1)}$, {\it lower}: $C^{(2)}$). Since we did not
detect any significant change in these coefficients for different mass
and redshift, we simply derive the values of $C^{(1)}$ and $C^{(2)}$ from
$\chi^2$ fits to the $N$-body data. We extrapolate the three coefficients to the
halo masses corresponding to the three surveys. The accuracy of these scalings can only be
tested using higher resolution simulations, and so leave this for future work.

\begin{figure}[htbp]
\begin{center}
\includegraphics[height=8cm]{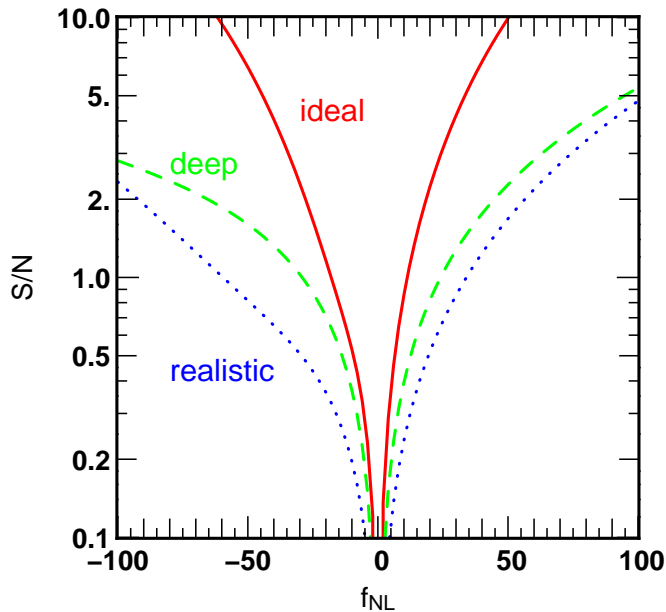}
\caption{Signal to noise ratios from the
  three future surveys defined in equation (\ref{eq:SN}). Notice this
  is estimated from only very limited configurations of Fourier space
  triangles: isosceles with two longer sides being
  $k_1=k_2=0.042h$Mpc$^{-1}$.}
\label{fig:SN}
\end{center}
\end{figure}

For statistical errors of these three surveys, we consider the
Gaussian contribution as a simple estimate, and neglect the
non-Gaussian error. We have \cite{Scoccimarro1998}:
\begin{eqnarray}
\hspace{-1cm}\left[\Delta\Bg(k_1,k_2,k_3)\right]^2 =
\frac{V}{N_{\rm triangle}}\left[P_g(k_1)+n_{\rm g}^{-1}\right]
\left[P_g(k_2)+n_{\rm g}^{-1}\right]\left[P_g(k_3)+n_{\rm g}^{-1}\right],
\label{eq:bis_error}
\end{eqnarray}
where $N_{\rm triangle}$ denotes the number of independent triangular
configurations in that bin, which roughly scales as $V^2$. We count
the number of triangles for each bin, $N_{\rm triangle}$, in
equation~(\ref{eq:bis_error}), assuming cubic-shaped survey with the
quoted volumes. See \ref{app:variance}, where we test this formula by comparison
with $N$-body simulations and show that it works reasonably well.
We use the linear power spectrum for $\Pg$ assuming the bias parameters
listed in Tab.~\ref{tab:survey}.

Now we are in a position to discuss about the future possibility to
detect the signature of the local-type primordial non-Gaussianity.  In
order to quantify the detectability from the $\alpha$-dependence of
the bispectrum, we define the signal-to-noise ratio:
\begin{eqnarray} 
    \left(\frac{\rm S}{\rm N}\right)^2 \equiv
    \sum_i\frac{\left[B(k,\alpha_i;\fnl)-B(k,\alpha_i;\fnl=0)\right]^2}
    {\Delta B(k,\alpha_i;\fnl=0)^2}. 
\label{eq:SN}
\end{eqnarray}
In evaluating Eq.~(\ref{eq:SN}), only the isosceles triangles with
$k=0.042h$Mpc$^{-1}$ are used, and the results are shown in the right hand panel
of Fig.~\ref{fig:SN}.  It is remarkable that even with the very
limited number of configurations for the bispectrum, detection of primordial
non-Gaussianity is possible in all three surveys if $\fnl$ is
several dozen. This can be compared with the analysis neglecting the
$B^{(2)}_{\rm g}$ term. According to \cite{Sefusatti2007}, using the
full configurations of the bispectrum leads to the constraint on
$\fnl$, $\Delta\fnl\simeq5-30$. Thus, we naively expected that using
the full configurations taking proper account of the scale dependence
of bispectrum greatly enhances the detectability of primordial
non-Gaussianity, and the constraints on $\fnl$ would be much tighter.

Finally, it is interesting to note that the $S/N$ for the deep survey
depends steeply on $\fnl$, and it exceeds that of the realistic survey 
of relatively large volume. This is primarily due to the $B^{(2)}_{\rm g}$ 
term, which scales as $\fnl^2$, being more significant at higher redshift, 
and thus helps to detect a small non-Gaussianity.  In this respect, the 
on-going mission BOSS \cite{BOSS}, aiming at precisely measuring the 
scale of baryon acoustic oscillations from the clustering of 
the LRGs at $z<0.7$ and QSO absorption systems at $z\sim2.5$, 
may be the promising probe for constraining or detecting primordial 
non-Gaussianity of local type.

\section{Conclusions and Discussion}
\label{sec:conclusion}
In this paper, we have studied the clustering properties of dark
matter haloes from cosmological $N$-body simulations in the presence
of local-type primordial non-Gaussianity. We found that the halo
bispectrum measured from $N$-body simulations exhibits a strong
$f_{\rm NL}$ dependence which becomes most prominent for squeezed
configurations at large scales. In particular, for realistic values of
$|\fnl|\simlt100$, the dependence of the halo bispectrum on $\fnl$ is well
characterized by the polynomial expansions of $\fnl$ up to second
order.  Since the quadratic dependence on $\fnl$ does not appear in
the matter bispectrum at the lowest order in perturbation theory, 
this would be a clear indicator for the existence of primordial non-Gaussianity of the local type.

We have investigated the shape and scale dependence of the halo
bispectrum arising from the $\fnl^2$ term, and the simulation results
are compared with theoretical predictions based on the local bias
model. For the isosceles triangles characterized by $\alpha\equiv k_1
/ k_3$ and $k\equiv k_1=k_2$, the dependence of the halo bispectrum on
$\alpha$ measured from $N$-body simulations is found to be consistent with
theoretical predictions by \cite{Jeong2009}.  We also examined the
dependence of the halo bispectrum on minimum halo mass and redshift, and
showed that the amplitude of the halo bispectrum is more significant for
more massive haloes at higher redshifts.

Thus, the strong dependence of the halo/galaxy bispectrum on $\alpha$ makes
the detection of primordial non-Gaussianity much more promising in
future surveys. As a preliminary investigation, we have evaluated the
signal-to-noise ratio for the scale dependence of the bispectrum in three
representative surveys, and found that even with the very limited number of 
configurations of bispectrum it is possible to detect primordial
non-Gaussianity if $\fnl$ is several dozen. Thus, the detectability
of primordial non-Gaussianity is expected to be greatly improved if we
use all configurations of the bispectrum.

We leave the following tasks as a future work: (i) Study the effects
of redshift-space distortions. Since we focus on very large scales, we
expect that these effects are accurately described by linear theory, 
i.e., they just enhance the
amplitude of the bispectrum in a scale independent way. (ii) Construct more
elaborate theoretical models that are applicable to a wider range of
triangular configurations and compare them with $N$-body simulations
(iii) Run higher resolution simulations where we can populate 
haloes with galaxies and measure the galaxy bispectrum directly from
simulations. These tasks are clearly very important to exploit future
surveys.

\ack We thank T.~Sousbie, R.~Nichol, E.~Komatsu, D.~Jeong, and Y.~Suto
for useful discussions and comments.  T.~N. is supported by a
Grant-in-Aid for Japan Society for the Promotion of Science (JSPS)
Fellows (DC1: 19-7066). A.~T. is supported by a Grant-in-Aid for
Scientific Research from JSPS (No. 21740168).  K.~K. is supported by
the European Research Council, Research Councils UK and the UK's
Science \& Technology Facilities Council (STFC).  C.~S. was funded by
a STFC PhD studentship.  Numerical computations were in part carried
out on Cray XT4 at Center for Computational Astrophysics, CfCA, of
National Astronomical Observatory of Japan.  We are also grateful for
the computational time provided by the U.K. National Grid Service
(NGS).  This work was supported in part by Grant-in-Aid for Scientific
Research on Priority Areas No.~467 ``Probing the Dark Energy through
an Extremely Wide and Deep Survey with Subaru Telescope'', and JSPS
Core-to-Core Program ``International Research Network for Dark
Energy''.
\appendix
\section{Power spectrum and bispectrum in the local bias model}
\label{app:pow_bis}
In this appendix, we compute the power spectrum and bispectrum of
biased tracers adopting the local bias model.

The local biasing scheme is a simple prescription to relate the
galaxy/halo density field, $\deltag$, to matter fluctuation,
$\deltam$, on large scales. In this treatment, the density fluctuation
of galaxies/haloes smoothed over the radius $R$, $\deltag$, is given
by a non-linear function of $\deltam$. On large scales, it can be
expanded as
\begin{equation}
\deltag(\bfx;R)=b_1 \,\deltam(\bfx;R) + \frac{b_2}{2} \,
\{\deltam^2(\bfx;R) -\sigmar^2\}+\cdots
\label{eq:local_bias}
\end{equation}
with $\sigmar$ being $\langle\deltam^2\rangle^{1/2}$. For simplicity,
we omit the dependence on redshifts throughout the
appendices. Equation (\ref{eq:local_bias}) can be rewritten in Fourier
space as
\begin{eqnarray}
\hspace{-2cm}\deltag(\bfk;R) &=& b_1\deltam(\bfk;R) \nonumber\\
\hspace{-3cm} && + \frac12b_2\int\frac{d^3\bfq}{(2\pi)^3}
\left[\deltam(\bfq;R)\deltam(\bfk-\bfq;R)-\langle\deltam(\bfq;R)\deltam(\bfk-\bfq;R)
\rangle\right].
\label{eq:bias_fourier}
\end{eqnarray}

Using this relation, let us consider the galaxy-matter cross spectrum. 
With the help of perturbative expansion, a straightforward calculation yields
\begin{eqnarray}
P_{\rm gm}(k;R) = b_1\Pm(k;R) + \frac12b_2\int\frac{d^3\bfq}{(2\pi)^3}
\Bm(q,k,|\bfk-\bfq|;R), 
\end{eqnarray}
which is valid up to third order in $\delta_{\rm m}$, and the
bispectrum $B_{\rm m}$ is given by the first term of equation
(\ref{eq:bis_pt}).  For the scales of our interest, the integration at
the right-hand side of this equation can be separately done taking the
large-scale limit.  We then obtain \cite{Taruya2008}
\begin{eqnarray}
P_{\rm gm}(k;R) &=& b(k,\fnl;R)\Pm(k;R)\,;
\\
b(k,\fnl;R)&=&b_1\left\{1+2\fnl\frac{b_2}{b_1}
\frac{\sigmar^2}{\mathcal{M}_R(k)}\right\}.
\end{eqnarray}
Here, we define 
\begin{eqnarray}
\mathcal{M}_R(k) \equiv \mathcal{M}(k)\tilde{W}_R(k),
\end{eqnarray}
with $\tilde{W}_R$ being the Fourier transform of the window function.

The above procedure can also be applied when we calculate the galaxy
auto power spectrum, $\Pg$. However, the derivation is rather
simplified if we recall that the deterministic bias relation holds on
large scales.  \footnote{This is indeed valid as long as we are
  concerned with the leading-order calculation. See \cite{Taruya2008}
  for alternative derivation.}  Then, the auto and cross power spectra
are tightly related with each other as $\{P_{\rm
  gm}(k;R)\}^2=\Pg(k;R)\,\Pm(k;R)$, which leads to
\begin{eqnarray}
\Pg(k;R) &=&  \Bigl\{b(k,\fnl;R)\Bigr\}^2\,\Pm(k;R)
\nonumber\\
&=&b_1^2\left\{1+2\fnl\frac{b_2}{b_1}\frac{\sigmar^2}
{\mathcal{M}_R(k)}\right\}^2\Pm(k;R).
\end{eqnarray}
On large scales, the effect of window functions is irrelevant, and we
simply drop the subscript $R$. Introducing the bias parameter
$\tilde{b}_2\equiv b_2\sigmar^2$, we finally obtain the galaxy power
spectrum {\it without smoothing}:
\begin{eqnarray}
\Pg(k) = b_1^2\left\{1+2\fnl\frac{\tilde{b}_2}{b_1}
\mathcal{M}^{-1}(k)\right\}^2\Pm(k),
\label{eq:pk}
\end{eqnarray}
which reproduces equation (\ref{eq:pow_bias}).  In the local bias
prescription, the bias parameters $b_1$ and $\tilde{b}_2$ are given
just as the fitting parameters. On the other hand, in the halo and
peak bias formalisms, these parameters have a specific functional
form, and are related with each other. We will discuss this issue in
\ref{app:bias}, and derive an explicit relation between $b_1$ and
$\tilde{b}_2$ [Eq.~(\ref{eq:relation_b1_b2})].

Next consider the galaxy bispectrum.  Again, starting from equation
(\ref{eq:bias_fourier}), a straightforward calculation yields
\cite{Jeong2009}
\begin{eqnarray}
\Bg(k_1,k_2,k_3;R) &=&
b_1^3
\left[
\Bm(k_1,k_2,k_3;R)+\frac{b_2}{b_1}
\left\{ \Pm(k_1;R)\Pm(k_2;R)+\mbox{(cyc.)}\right\}\right.\nonumber\\
&&\left.+\frac{\tilde{b}_2}{b_1}B_{\rm corr}(k_1,k_2,k_3;R)\right], 
\label{eq:Bispectrum_JK09}
\end{eqnarray}
which is valid up to fourth order in $\delta_{\rm m}$. In the
above, the quantities $\Pm$ and $\Bm$ are the matter power spectrum
and bispectrum, whose perturbative expressions are given in equations
(\ref{eq:pow_pt}) and (\ref{eq:bis_pt}).  On the other hand, the term
$B_{\rm corr}$ represents a new contribution arising from the matter
trispectrum, $\Tm$, and it is expressed as
\begin{eqnarray}
    B_{\rm corr}(k_1,k_2,k_3;R) &\equiv&
    \frac{1}{2\sigmar^2}\int\frac{d^3\bfq}{(2\pi)^3}
    \left[\Tm(\bfq,\bfk_1-\bfq,\bfk_2,\bfk_3;R)+\mbox{(cyc.)}\right]. 
\end{eqnarray}
According to the perturbative calculation by \cite{Jeong2009},  the 
above equation can be further decomposed into several pieces as: 
\begin{eqnarray}
    B_{\rm corr}(k_1,k_2,k_3;R) &=& \fnl^2\,
    B_{\fnl^2}^{\rm nG}(k_1,k_2,k_3;R)
    \nonumber\\
    &&
    + \fnl\,\left[\Bm^{\rm nG}(k_1,k_2,k_3;R) +
        B_{\fnl}^{\rm nG1}(k_1,k_2,k_3;R)\right.\nonumber\\
    &&\left.+4B_{\fnl}^{\rm nG0}(k_1,k_2,k_3;R)\sum_{i=1}^3 \mathcal{G}_{\rm R}(k_i)
    \right],
\end{eqnarray}
Here, the term $B_{\fnl^2}^{\rm nG}$ comes from the leading-order
trispectrum.  For $k\simlt0.1h$ Mpc$^{-1}$, it is approximated as
\begin{eqnarray}
B_{\fnl^2}^{\rm nG}(k_1,k_2,k_3;R)
&\approx &
\frac{1}{2\sigmar^2}\biggl[
8\Mr(k_2)\Mr(k_3)
P_\phi(k_1)
\left[
P_\phi(k_2)+P_\phi(k_3)
\right]\nonumber\\
&&\times\int \frac{d^3\bfq}{(2\pi)^3}
\Mr(q)
\Mr(|\bfk_1-\bfq|)
P_\phi(q)
+
\cyc
\nonumber\\
&&+
4\Mr(k_2)\Mr(k_3)P_\phi(k_2)P_\phi(k_3)\nonumber\\
&&
\int \frac{d^3\bfq}{(2\pi)^3}
\Mr(q)
\Mr(|\bfk_1-\bfq|)
\nonumber\\
& &\times
\left[
P_\phi(|\bfk_2+\bfq|)
+
P_\phi(|\bfk_3+\bfq|)
\right]
+
\cyc\biggr]
\end{eqnarray}
with $P_\phi$ being the power spectrum of $\Phi_{\rm G}$.  The
explicit expressions for the other remaining terms are obtained by
integrating the next-to-leading order contributions to the
trispectrum. The resultant expressions become
\begin{eqnarray}
\Bm^{\rm nG}(k_1,k_2,k_3;R)
&=&
4\Wr(k_1)\Wr(k_2)\Wr(k_3)
\left[
\frac{\Fr(k_1)}{\Mr(k_1)}
+
\frac{\Fr(k_2)}{\Mr(k_2)}
\right] \nonumber\\
&&\times\Pm(k_1)\Pm(k_2)F_2^{\rm (s)}(\bfk_1,\bfk_2)
+\cyc,\\
B_{\fnl}^{\rm nG1}(k_1,k_2,k_3;R)
&\approx&
\frac{1}{2\sigmar^2}
\biggl[
8\Wr(k_2)\Wr(k_3)
P_m(k_2)\Mm(k_3)P_\phi(k_3)
\nonumber\\
&&\times \int \frac{d^3q}{(2\pi)^3}
\Wr(|\bfk_1-\bfq|)
\Wr(q)
\Mm(|\bfk_1-\bfq|)
\Mm(|\bfk_2+\bfq|)
\nonumber\\
&&\times
\left[
P_\phi(|\bfk_2+\bfq|) +
P_\phi(|\bfk_1-\bfq|)
\right]
F_2^{(s)}(-\bfk_2,\bfk_2+\bfq)
\nonumber\\
&&+(5~\mathrm{permutation})\bigg]. 
\end{eqnarray}
The term $B^{\rm nG0}_{\fnl}$ coincides with the first term of the
matter bispectrum in equation (\ref{eq:bis_pt}).  The functions ${\cal
  F}_{\rm R}(k_i)$ and ${\cal G}_{\rm R}(k_i)$ weakly depend on the
smoothing scale $R$.  Note that in deriving the above expressions, we
have neglected the irrelevant terms at $k\simlt0.1h$ Mpc$^{-1}$.

In the expression for the galaxy bispectrum, the important findings
here are a new term which scales as $\fnl^2$ and additional
contributions which scale as $\fnl$ to the matter bispectrum. Although
\cite{Jeong2009} further considered the term arising from a cubic
correction, $\gnl\Phi_{\rm G}^3$ to equation (\ref{eq:fnllocal}), we
do not discuss about this term in this paper. See also
\cite{Desjacques2009b,Giannantonio2009,Sefusatti2009} for discussions
about the $\gnl$ term.  Ref. \cite{Jeong2009} evaluates the asymptotic
forms of each term at squeezed limit ($\alpha\gg1$, where
$k_1=k_2=\alpha k_3=k$) for isosceles triangular configurations, and
the results are shown in equation (\ref{eq:Bsqueeze}) in the text.
Note that Ref.~\cite{Sefusatti2009} also compute the matter and galaxy
bispectrum up to the one-loop order (i.e., $\mathcal{O}(\delta_0^5)$)
assuming local bias model for both local and equilateral type
non-Gaussianity.

Alternatively we might be able to investigate the galaxy bispectrum
using a similar description to the galaxy power spectrum:
\begin{eqnarray}
\hspace{-1.5cm}\Bg(\bfk_1,\bfk_2,\bfk_3;R)&=&b(k_1,\fnl;R)\,b(k_2,\fnl;R)\,b(k_3,\fnl;R)\,
\Bm(\bfk_1,\bfk_2,\bfk_3;R).
\label{eq:bg}
\end{eqnarray}
We focus on squeezed triangles at large scales, $k\to0$ and
$\alpha\gg1$ and drop the window function. Then we get the
second-order term in $\fnl$, which is the same as the term in equation
(\ref{eq:B2}) up to the prefactor. However, the coefficient of the
term scaling as $\fnl$ in this prescription is
$4b_1^3+(26/7)b_1^2\tilde{b}_2$ in this limit, which does not
reproduce the contribution that depends on the smoothing scale in
equation (\ref{eq:B1}). This prescription also misses the term coming
from $b_1^2b_2[\Pm(k_1)\Pm(k_2) +\cyc]$.  This implies that although
some contributions are not included, this description captures the
essence of the galaxy bispectrum at the squeezed limit.  In other
words, the $\fnl^2$ term has the same origin as the scale dependent
bias in the power spectrum. Although equation~(\ref{eq:bg}) seems rather
empirical, it is naturally derived in Ref.~\cite{Giannantonio2009}, where the authors
calculated the galaxy bispectrum using a multivariate biasing scheme [see their equations~(73) - (76)].

\section{On the relation between peak biasing and
  local biasing models}
\label{app:bias}

Scale-dependence of halo power spectrum and bispectrum 
in the presence of primordial non-Gaussianity have been derived in the 
literature in different ways, 
based on the local bias prescription and the halo/peak formalism. 
However, the resultant expressions for peak and halo bias 
coincide with each other in the high-peak/thresold limit, and 
there is a clear relationship between peak bias and local bias 
prescriptions. The relation between the linear and quadratic 
bias parameter plays important roles in the accurate modelling for
the power spectrum and the bispectrum of galaxies as
seen in the previous appendix.

In this appendix, in order to elucidate these 
properties in a self-contained manner, we give an explicit
relationship between the local biasing and peak biasing models, and 
show that the peak density field in the high-peak limit can be
described by the local biasing prescription. In the end, we obtain 
the relation (\ref{eq:relation_b1_b2}), which was used in 
Sec.~\ref{sec:pow_results} when comparing the $N$-body results with 
model prediction of halo power spectrum. We also use this relation
in Sec.~\ref{sec:discussion} to discuss the future detectability of the
local-type primordial non-Gaussianity.


We begin by writing down the definition of peak density field. 
According to \cite{Matarrese1986}, it is given by 
\begin{equation}
\deltag(\bfx;R)=\frac{\rho_\nu(\bfx;R)}{\langle\rho_\nu(\bfx;R)\rangle }-1;
\quad
\rho_\nu(\bfx;R)\equiv\Theta\Bigl[\deltam(\bfx;R)-\nu\,\sigmar\Bigr],
\label{eq:def_peak}
\end{equation}
in the Lagrangian space. In the above, $\Theta$ is the Heaviside step
function and $\nu\equiv\deltac/\sigmar$ with the critical overdensity,
$\deltac\simeq1.686$. Strictly speaking, the above definition does not
imply the local maximum of the density field, however, the local
density specified above is expected to roughly correspond to the peak
in the high-threshold limit, $\nu\gg1$.

Starting with the expression (\ref{eq:def_peak}), we want to derive 
Taylor series expansion of $\deltag$ in terms of the 
local density $\deltam$. 
To do this, let us first expand the peak density in terms of the Hermite 
polynomials:
\begin{equation}
\rho_\nu(\bfx;R)=\sum_{n=0}^{\infty} \frac{R_n}{n!} \,H_n(\deltam/\sigmar).
\label{eq:rho_peak}
\end{equation}
The coefficient $R_n$ is given by (e.g., \cite{Matsubara1995}):
\begin{equation}
\hspace{-1cm}R_n =\int_{-\infty}^{+\infty} \frac{dy}{\sqrt{2\pi}}\,e^{-y^2/2} H_n(y)\,
\Theta\bigl[(y-\nu)\,\sigmar\bigr]
=\left\{
\begin{array}{ccl}
{\displaystyle \frac{1}{2}\,\mbox{erfc}\left(\frac{\nu}{\sqrt{2}}\right)}
&;& n=0,
\\
\\
{\displaystyle \frac{e^{-\nu^2/2}}{\sqrt{2\pi}}H_{n-1}(\nu)} &;& n\geq1.
\end{array}
\right.
\end{equation}
In the high-peak limit $\nu\gg1$, the coefficient $R_n$ asymptotically
approaches
\begin{equation}
R_n\,\, \longrightarrow \,\, \frac{\nu^{n-1}}{\sqrt{2\pi}}\,e^{-\nu^2/2},
\end{equation}
for $n\geq0$. Substituting this back into (\ref{eq:rho_peak}), we obtain
\begin{equation}
\rho_\nu(\bfx;R)\simeq
\frac{e^{-\nu^2/2}}{\nu\sqrt{2\pi}}\,
\sum_{n=0}^{\infty} \,
\frac{\nu^n}{n!}\,H_n(\deltam/\sigmar)
=\frac{1}{\nu\sqrt{2\pi}}\,e^{-\nu^2+\nu(\deltam/\sigmar)},
\end{equation}
where we used the relation $\sum_n (x^n/n!)H_n(t)=e^{-x^2/2+t\,x}$
in the last equality. Now, recalling
from the cumulant expansion theorem,
$\langle e^{it\deltam}\rangle=\exp[\sum_n\,(it)^n\langle\deltam^n\rangle_c/n!]$,
the averaged peak density $\langle\rho_\nu\rangle$ in the high-peak
limit becomes
\begin{equation}
\hspace{-1.5cm}\bigl\langle\rho_\nu(\bfx;R)\bigr\rangle \simeq
\frac{e^{-\nu^2}}{\nu\sqrt{2\pi}}
\,\,\bigl\langle e^{(\nu/\sigmar)\,\deltam}\bigr\rangle
= \frac{e^{-\nu^2}}{\nu\sqrt{2\pi}} \,\,\exp\left[\sum_{n=0}^\infty \,
\frac{(\nu/\sigmar)^n}{n!}\,
\Bigl\langle\{\deltam(\bfx;R)\}^n\Bigr\rangle_c\right].
\end{equation}
Hence, the peak density field $\deltag$ becomes
\begin{equation}
  \label{eq:delta_nu}
  \deltag(\bfx;R)\simeq \exp\left[\frac{\nu}{\sigmar}\,\deltam(\bfx;R)
-\sum_{n=0}^\infty \,\frac{(\nu/\sigmar)^n}{n!}\,
\Bigl\langle\{\deltam(\bfx;R)\}^n\Bigr\rangle_c\right]-1,
\end{equation}
which can be expanded in the form of local biasing expression
(\ref{eq:local_bias}) as
\begin{equation}
\hspace{-2cm}  \deltag(\bfx;R)=\frac{\nu}{\sigmar}\,\deltam(\bfx;R)+
\frac{1}{2}\left(\frac{\nu}{\sigmar}\right)^2
\{\deltam^2(\bfx;R) -\sigmar^2\}+
\frac{1}{3!}\left(\frac{\nu}{\sigmar}\right)^3
\{\deltam^3(\bfx;R) -\langle\deltam^3\rangle_c\}+\cdots.
\label{eq:peak_bias_expand}
\end{equation}

Note that this expansion is done in Lagrangian space. Assuming the usual
mapping from Lagrangian to Eulerian space, $b_{\rm E}=1+b_{\rm L}$,
where $b_{\rm E}$ and $b_{\rm L}$ are linear bias parameters in
Eulerian and Lagrangian space, the biasing parameters in the high-peak
limit can be read off by comparing equation
(\ref{eq:peak_bias_expand}) with equation (\ref{eq:local_bias}), and
are expressed as
\begin{equation}
  b_1=1+\frac{\nu^2}{\deltac},\quad
  b_2=\frac{\nu^4}{\deltac^2},\quad
  b_3=\frac{\nu^6}{\deltac^3}, \quad\cdots .
\label{eq:b_peak}
\end{equation}
Note that the higher-order biasing parameters should be also modified
by the mapping from Lagrangian and Eulerian space but this effect is
small in the high peaks limit and we simply ignore it. 
Then the biasing parameters have a relation
\begin{eqnarray}
\tilde{b}_2 = \deltac(b_1-1),
\label{eq:relation_b1_b2}
\end{eqnarray}
where $\tilde{b}_2 = b_2 \sigma_{\rm R}^2$. 
For a better fit of halo power spectrum (\ref{eq:pow_bias}) to 
the N-body simulations,  a slight modification to the above relation 
might be necessary [see Eq.~(\ref{eq:q_grossi})]. 

\section{The Sample Variance of the Matter and Halo Bispectrum}
\label{app:variance}
It is of importance to investigate the variance of the bispectrum in the presence
of primordial non-Gaussianity, although our simulation sets are too small to examine 
the full covariance of the bispectrum (see e.g., \cite{Takahashi2009}; the authors
performed $5000$ realizations of $(1h^{-1}$Gpc$)^3$ volume simulations to investigate
it for Gaussian initial conditions).
Here we show our measurements of the variance (i.e., the diagonal elements of the
covariance matrix) for both the matter and halo bispectrum.

\begin{figure}[htbp]
\begin{center}
\includegraphics[height=6.5cm]{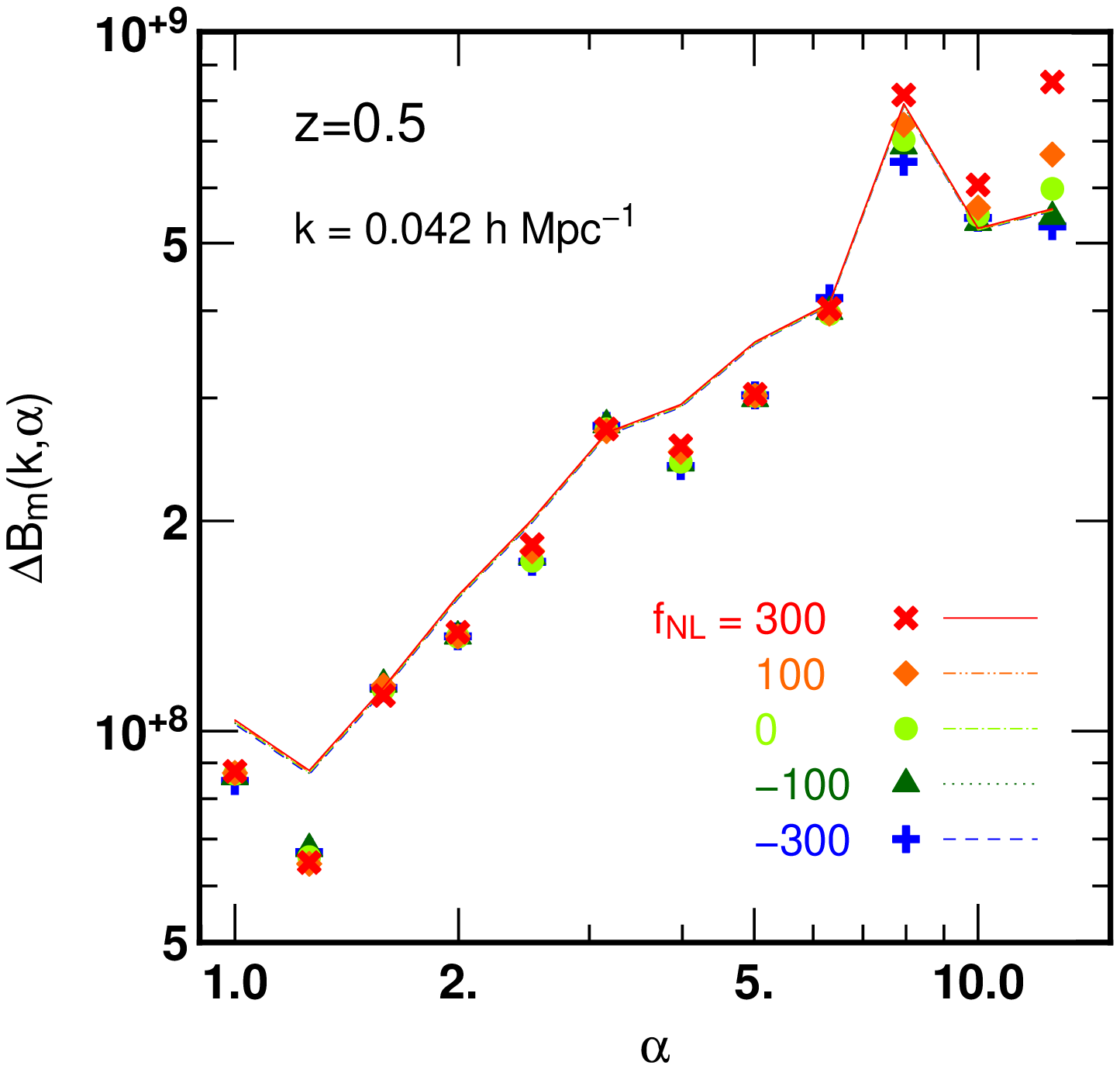}
\includegraphics[height=6.5cm]{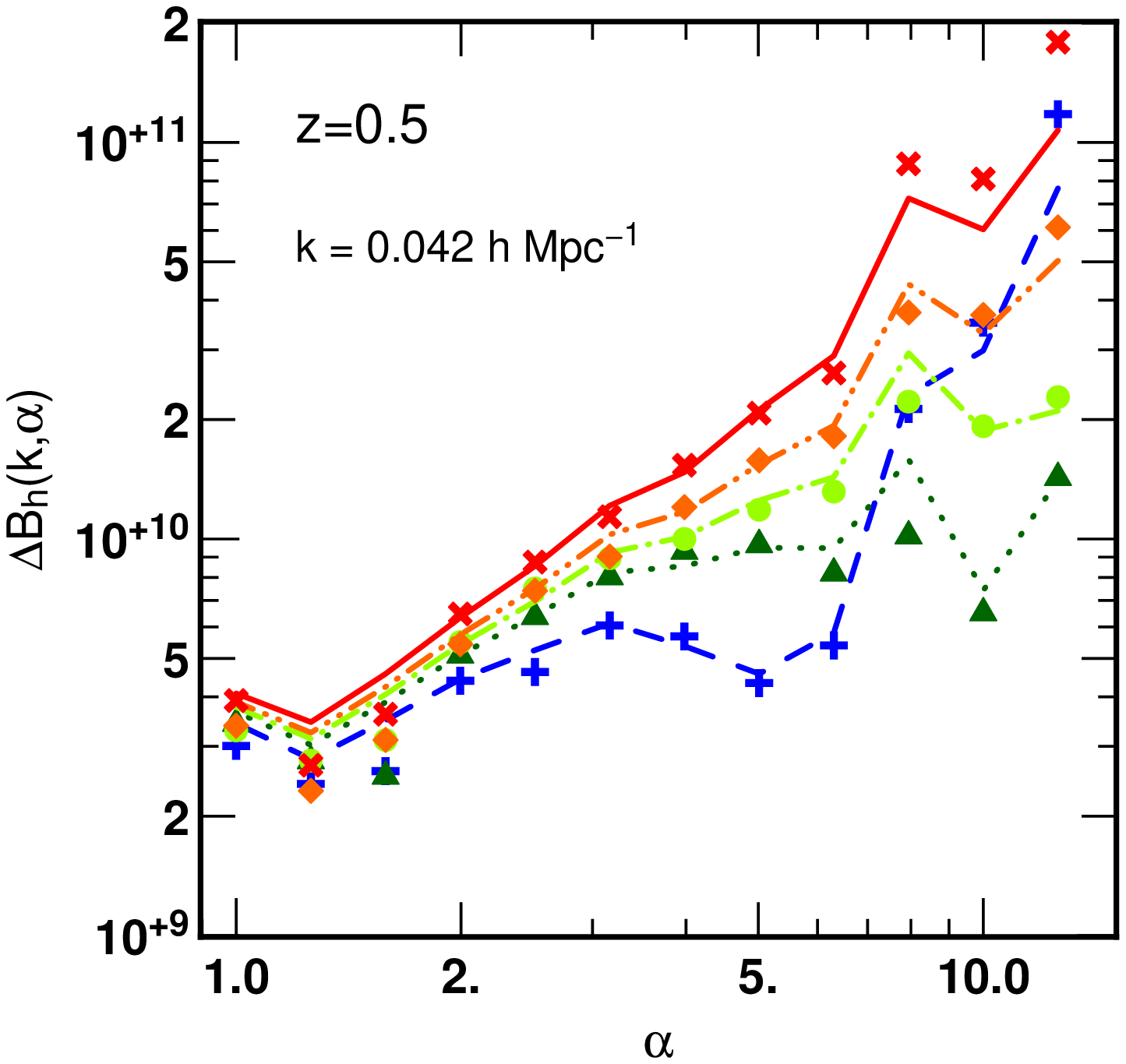}
\caption{Variance of the matter (left) and halo (right)
  bispectrum. The symbols are measured from $N$-body simulations,
  while the lines are computed by equation~(\ref{eq:bis_error}).}
\label{fig:bis_error}
\end{center}
\end{figure}

Fig.~\ref{fig:bis_error} shows the variance of the matter (left) and halo (right) bispectrum for the 
same configurations as in the left panel of FIg.~\ref{fig:dependence} at $z=0.5$. 
The symbols correspond to the variance measured from $N$-body simulations, while the lines
are obtained from equation~(\ref{eq:bis_error}). In computing equation~(\ref{eq:bis_error}),
we substitute the value of the power spectrum measured from $N$-body simulations both 
for matter and halo. Overall, the analytic predictions are good approximation of the $N$-body 
simulations, ensuring the use of this formula. For the variance of the matter bispectrum,
there is little evidence of $\fnl$ dependence. This is natural because the matter power spectrum
also depends on $\fnl$ only weakly (see Fig.~\ref{fig:dmpow}). On the other hand, the right
panel shows a strong dependence on $\fnl$, reflecting a strong $\fnl$ dependence of the halo power
spectrum. The prediction of equation~(\ref{eq:bis_error}) seems worse at larger $\alpha$ and
larger $|\fnl|$. Since the leading correction term to this formula have the form of 
$P(k_1)T(\bfk_2,\bfk_3,-\bfk_2,-\bfk_3)+\cyc$, this feature is quite reasonable.



\section*{References}
\begin{thebibliography}{99}
\bibitem{WMAP5}
  Komatsu,~E. \etal, 2008, ApJS, 180, 330
\bibitem{Tegmark2006}
  Tegmark,~M. et al. 2006, Phys.Rev.D 69, 123507
\bibitem{Komatsu2001}
  Komatsu,~E., \& Spergel,~D.~N., 2001, PRD 63, 063002
\bibitem{Carbone2008}
  Carbone,~C., Verde,~L., \& Matarrese,~S., 2008, ApJ, 684, 1
\bibitem{Bartolo2004}
  Bartolo,~N., Komatsu,~E., Matarrese,~S., \& Riotto,~A., 2004,
  Phys. Rep. 402, 103
\bibitem{Planck}
 [Planck Collaboration], arXiv:astro-ph/0604069
\bibitem{Scoccimarro2000}
  Scoccimarro,~R., 2000, ApJ. 542, 1
\bibitem{Verde2000}
  Verde,~L., Wang,~L., Heavens,~A.~F., \& Kamionkowski,~M., 2000, MNRAS, 313, 141
\bibitem{Scoccimarro2004}
  Scoccimarro,~R.,  Sefusatti,~E., \& Zaldarriaga,~M., 2004, PRD, 69, 103513
\bibitem{Sefusatti2007}
  Sefusatti,~E., and Komatsu,~E., 2007, PRD, 76, 083004
\bibitem{Dalal2008}
  Dalal,~N., Dor\'e,~O., Huterer,~D., and Shirokov,~A., 2008, PRD, 77, 12351
\bibitem{Afshordi2008}
  Afshordi,~N., and Tolley,~A.~J., 2008, PRD, 78, 123507
\bibitem{Taruya2008}
  Taruya,~A., Koyama,~K., \& Matsubara,~T., 2008, PRD, 78, 123534
\bibitem{Matarrese2008}
  Matarrese,~S., and Verde,~L., 2008, AJ, 677, L77
\bibitem{Pillepich2008}
  Pillepich,~A., Porciani,~C., and Hahn,~O., 2010, MNRAS, 402, 191
\bibitem{Desjacques2009a}
  Desjacques,~V., Seljak,~U., \& Iliev,~I.~T., 2009, MNRAS, 396, 85
\bibitem{Desjacques2009b}
  Desjacques,~V., \& Seljak,~U., 2010, PRD, 81, 023006
\bibitem{Grossi2009}
  Grossi,~M., \etal, 2009, MNRAS, 398, 321
\bibitem{Slosar2008}
  Slosar,~A., Hirata,~C., Seljak,~U., Ho,~S., and Padmanabhan,~N.,
  2008, JCAP, 08, 031
\bibitem{McDonald2008}
  McDonald,~P., 2008, PRD, 78, 123519
\bibitem{Sefusatti2009}
  Sefusatti,~E., 2009, PRD, 80, 123002
\bibitem{Giannantonio2009}
  Giannantonio,~T., \& Porciani,~C., 2010, PRD, 81, 063530
\bibitem{Jeong2009}
  Jeong,~D, \& Komatsu,~E., 2009, ApJ, 703, 1230
\bibitem{Matarrese1986}
  Matarrese,~S., Lucchin,~F., \& Bonometto,~S.~A., 1986, ApJ, 310, 21
\bibitem{Fry1993}
  Fry,~J.~N., and Gaztanaga,~E., 1993, ApJ, 413, 447
\bibitem{Bernardeau2002}
  Bernardeau,~F., Colombi,~S., Gaztan\~aga,~E., and Scoccimarro,~R., 2002,
  Phys. Rep., 367, 1
\bibitem{CAMB}
  Lewis,~A. \etal, 2000, AJ, 538, 473
\bibitem{GADGET2}
  Springel,~V., 2005, MNRAS, 364, 1105
\bibitem{Taruya2009}
  Taruya,~A., Nishimichi,~T., Saito,~S., \& Hiramatsu,~T., 2009, PRD, 80, 123503
\bibitem{Nishimichi2009}
  Nishimichi,~T. \etal, 2009, PASJ, 61, 321
\bibitem{Crocce2006}
  Crocce,~M., Pueblas,~S., and Scoccimarro,~R., 2006, MNRAS, 373, 369
\bibitem{LoVerde2008}
  LoVerde,~M., Miller,~A., Shandera,~S., and Verde,~L.,
  2008, JCAP, 04, 014
\bibitem{Matarrese2000}
  Matarrese,~S., Verde,~L., \& Jimenez,~R., 2000, ApJ, 541, 10
\bibitem{Warren2006}
  Warren,~M.~S., Abazajian,~K., Holz,~D.~E., Teodoro,~L., 2006, ApJ, 646, 881
\bibitem{Hockney1981}
  Hockney,~R.~W, \& Eastwood,~J.~W., 1981, Computer Simulations Using Particles
  (New York: McGraw-Hill)
\bibitem{Bartolo2010}
  Bartolo,~N., Beltr\'{a}n Almeida,~J.~P., Matarrese,~S., Pietroni,~M., 
  \& Riotto,~A., 2010 JCAP, 03, 011
\bibitem{EUCLID}
   http://www.ias.u-psud.fr/imEuclid/
\bibitem{SuMIRe}
  Aihara,~T., talk at the IPMU international conference on dark energy:
  lighting up the darkness!
\bibitem{HETDEX}
   http://www.as.utexas.edu/hetdex/
\bibitem{Crocce2009}
  Crocce,~M., Fosalba,~P., Castander,~F.~J., \& Gaztan\~aga,~E., 2010, MNRAS, 403, 1353
\bibitem{Sheth1999}
  Sheth,~R.~K., Tormen,~G., 1999, MNRAS, 308, 119
\bibitem{Scoccimarro1998}
  Scoccimarro,~R., Colombi,~S., Fry,~J.~N., Frieman,~J.~A., Hivon,~Eric, \&
  Melott,~A, 1998, ApJ, 496, 586
\bibitem{BOSS}
  http://cosmology.lbl.gov/BOSS/
\bibitem{Matsubara1995}
  Matsubara,~T., 1995, APJS, 101, 1
\bibitem{Takahashi2009}
  Takahashi.~R., et al., 2009, ApJ, 700, 479
\end{thebibliography}
\end{document}